\documentclass[%
 reprint,
superscriptaddress,
nofootinbib,
 amsmath,amssymb,
]{revtex4-2}

\usepackage[utf8]{inputenc}
\usepackage{graphicx}
\usepackage{bm}
\usepackage{siunitx}
\usepackage[left=2.5cm, right=2.5cm, top=2.5cm]{geometry}
\usepackage{amsmath,amssymb}
\usepackage{color}
\usepackage{comment}
\usepackage{xcolor}
\usepackage{float}
\usepackage[colorlinks=true,allcolors=blue]{hyperref}%

\usepackage{booktabs} 

\newcommand{\ket}[1]{|#1\rangle}
\newcommand{\braket}[1]{\langle #1 \rangle}
\newcommand{\gp}[1]{{\color{black}#1}}

\newcommand{\crr}[1]{{\color{black}#1}}

\begin{document}
\preprint{APS/123-QED}

\title{\crr{
Delocalized Excitation Transfer in Open Quantum Systems \\ with Long-Range Interactions
}}
\author{Diego Fallas Padilla}
\email{difa1788@colorado.edu}
\affiliation{Department of Physics and Astronomy and Smalley-Curl Institute, Rice University, Houston, TX 77005, USA}
\affiliation{JILA and Department of Physics, University of Colorado, 440 UCB, Boulder, CO 80309, USA}
\affiliation{Center for Theory of Quantum Matter, University of Colorado, Boulder, CO 80309, USA}
\author{Visal So}
\affiliation{Department of Physics and Astronomy and Smalley-Curl Institute, Rice University, Houston, TX 77005, USA}
\author{Abhishek Menon}
\affiliation{Department of Physics and Astronomy and Smalley-Curl Institute, Rice University, Houston, TX 77005, USA}
\author{Roman Zhuravel}
\affiliation{Department of Physics and Astronomy and Smalley-Curl Institute, Rice University, Houston, TX 77005, USA}
\author{Han Pu}
\affiliation{Department of Physics and Astronomy and Smalley-Curl Institute, Rice University, Houston, TX 77005, USA}
\author{Guido Pagano}
\email{pagano@rice.edu}
\affiliation{Department of Physics and Astronomy and Smalley-Curl Institute, Rice University, Houston, TX 77005, USA}

\begin{abstract}


The interplay between coherence and system-environment interactions is at the basis of a wide range of phenomena, from quantum information processing to charge and energy transfer in molecular systems, biomolecules, and photochemical materials. In this work, we use a Frenkel exciton model with long-range interacting qubits coupled to a damped collective bosonic mode to investigate vibrationally assisted transfer processes in donor-acceptor systems featuring internal substructures analogous to light-harvesting complexes. We find that certain delocalized excitonic states maximize the transfer rate and that the entanglement is preserved during the dissipative transfer over a wide range of parameters. We investigate the reduction in transfer caused by static disorder, \crr{white noise}, and finite temperature and study how transfer efficiency scales as a function of the number of dimerized monomers and the component number of each monomer, finding which excitonic states lead to optimal transfer. Finally, we provide a realistic experimental setting to realize this model in analog trapped-ion quantum simulators. Analog quantum simulation of systems comprising many and increasingly complex monomers could offer valuable insights into the design of light-harvesting materials, particularly in the non-perturbative intermediate parameter regime examined in this study, where classical simulation methods are resource-intensive.


\end{abstract}

\maketitle

\section{Introduction}\label{sec_intro}


Understanding how quantum systems interact with their surrounding environment and how coherence and quantum correlations affect the non-equilibrium dynamics in the presence of decoherence and dissipation remains one of the outstanding challenges in quantum information science, condensed matter physics, and physical chemistry. The delicate interplay of dissipative and coherent processes in driven-dissipative quantum systems can lead to non-trivial out-of-equilibrium dynamics \cite{fazio2024manybodyopenquantumsystems}, dissipative phase transitions that do not always have an equilibrium counterpart \cite{Lee2013,sierant2022dissipative, debecker2024controlling,debecker2024spectral}, and dissipative engineering of correlated many-body steady states \cite{diehl2008quantum,Verstraete2009,cho2011,Huelga2012, Gonzalez2013, Kienzler2015, Morigi2015, Harrington2022, Cole2022, PineiroOrioli2022, Sundar2024}. 

Open quantum systems are also used to model the transfer of excitations coupled to nuclear vibrations in the presence of a complex environment, which occurs in charge transfer (CT) and energy transfer (ET) phenomena in various systems in molecular electronics \cite{ratner2013brief}, biomolecules \cite{Wang2019quantum, Scholes2017}, Perovskite photochemical cells \cite{gallop2024ultrafast}, and bi-layer materials \cite{park2021coherent}. 
In these systems, environmental fluctuations and vibrations play a crucial role in assisting the transfer of excitations to the lowest-energy state by dissipating energy and destroying coherent superpositions, ensuring irreversible transfer \cite{Monahan2015, Fujihashi2015}. For example, the role of vibronic coherence \gp{and delocalization} has been suggested to be connected to the fast transfer of excitons in light-harvesting complexes \cite{Engel2007, Romero2014, Fassioli2014, Arsenault2020} \gp{and organic semiconductors \cite{Sneyd2021, Sneyd2022}}. 

CT and ET in the presence of a complex environment can be generally described by Frenkel exciton models \cite{Wolynes2009, Ishizaki2009, Plenio2013, Jang2018} that employ discrete Hilbert spaces to encode the electronic states along with harmonic oscillators to describe molecular and solvent vibrations. Recently, it has been shown that both electronic and vibrational degrees of freedom can be directly mapped onto state-of-the-art analog quantum simulators based on trapped atomic ions \cite{Kang2024} and superconducting qubits \cite{Dutta2024}. These platforms allow for precise control of the coherent part of the evolution \cite{Gorman2018, Sun2023, whitlow2023, Valahu2023, cabral2024roadmap, dealbornoz2024oscillatory, Frattini2024} and, in some cases, also provide turning knobs to tailor the system-bath coupling implementing reservoir engineering \cite{Potonik2018, so2024trappedion, Wang2024simulating, sun2024quantumsimulation}.


In this work, we theoretically and numerically investigate a minimal Frenkel exciton model, where the donor and acceptor have an internal substructure \cite{Mattioni2021} similar to light-harvesting complexes in which different monomers feature strongly coupled molecular sites. Here, the molecular sites are encoded in collections of two-level systems, among which the excitation can be shared. The two-level systems are coupled to a damped collective bosonic mode, where the damping arises from the presence of an Ohmic bath.

\begin{figure}[h!]
\includegraphics[width=0.45\textwidth]{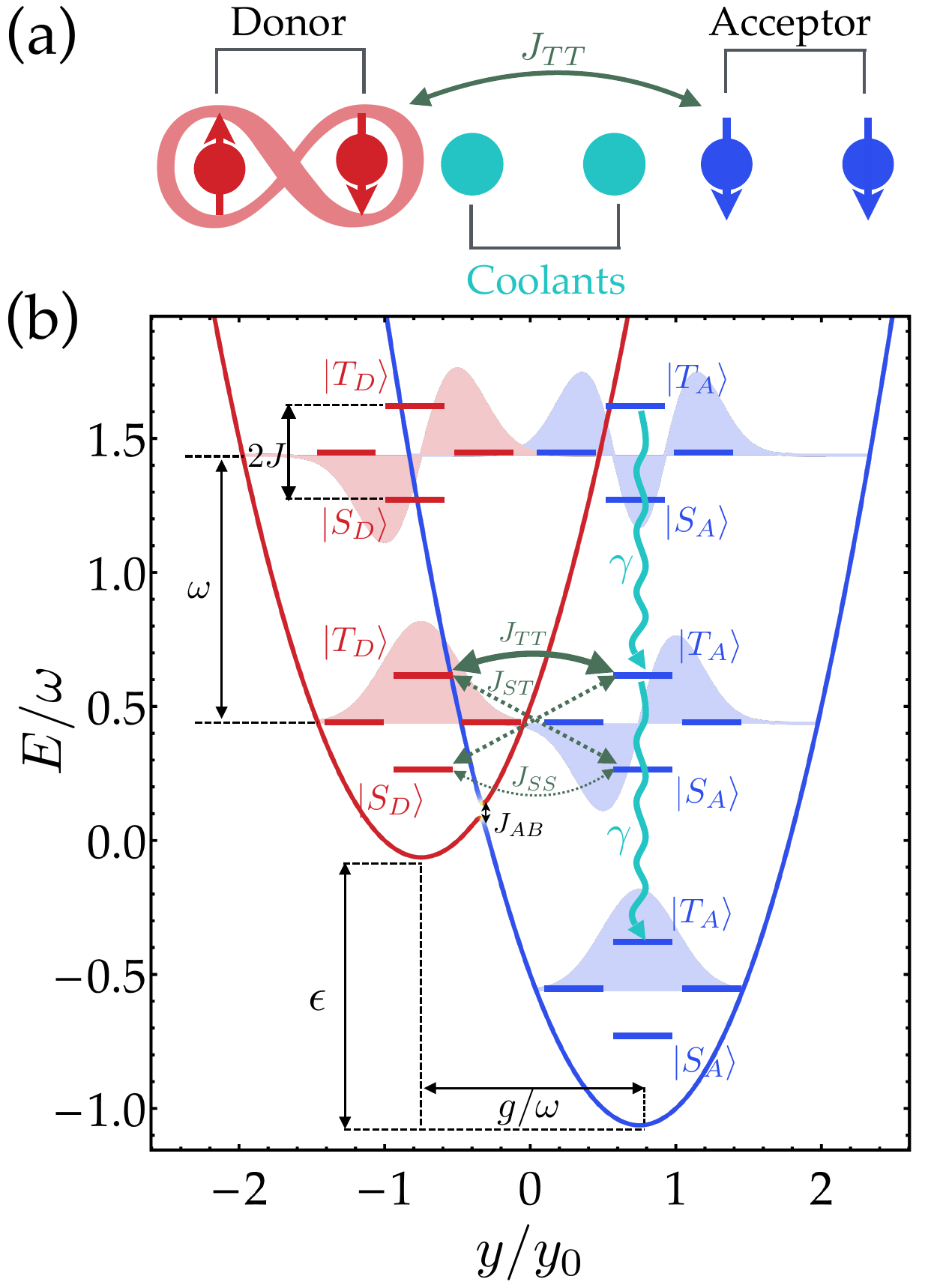}
\caption{{\bf Model and trapped-ion implementation:} 
{\bf (a)} A chain of 4 qubit ions and 2 coolants can realize the Frenkel exciton model (see main text). The red (blue) ions on the sides represent donor and acceptor monomers, respectively, and the information in each ion is encoded in a qubit. The two cyan ions in the middle represent the cooling ions used for the sympathetic cooling of the ion chain motion. 
{\bf (b)} Pictorial representation of the spectrum of the total Hamiltonian in the symmetric case, described by Eqs.~\eqref{H0J} and \eqref{HperTS}.
Donor (red) and acceptor (blue) surfaces are shown as a function of the normalized reaction coordinate $y/y_0=1/2(a^\dag+a)$ with their respective non-interacting harmonic wavefunctions. The effect of the bath is illustrated by the cyan wavy arrows that represent the relaxation with rate $\gamma$. 
The qubits in the donor/acceptor sites are strongly coupled and can be described in terms of local triplet and singlet states $\ket{T_{D,A}},\ket{S_{D,A}}$ separated by an energy $2J$. These states are, in turn, coupled by inter-monomer couplings $J_{AB}$, where $AB=TT,ST,SS$ (see main text). The inter-monomer couplings $J_{AB}$ induce anti-crossings
between the two surfaces allowing the donor to acceptor transfer. }
\label{Fig1}
\end{figure}

Previous studies focused on the incoherent transfer among weakly coupled monomers \cite{Mattioni2021} or coherent exciton transfer with nearest-neighbor couplings \cite{Myers2015vaenergypumps}.
Conversely, in this work, we study a regime in which inter-monomer coherences play a key role in the excitation transfer, and the molecular sites possess long-range interactions. The model studied here singles out the role of delocalization in excitation transfer and its robustness against static disorder in the energy landscape and the vibronic couplings, as well as the effect of the temperature. We also generalize this setting for a larger number of monomers and for monomers with increasingly complex internal substructures.

In addition to its fundamental theoretical interest, this model is also motivated by being experimentally realizable on an analog trapped-ion simulator, where the electronic degrees of freedom can be encoded in the internal atomic states, and the molecular vibrations in the collective vibrations of the ions around their equilibrium positions (see Section \ref{sec_ions}). Finally, in this work, we provide experimentally accessible parameters to realize this model in a trapped-ion simulator using a chain of qubit and coolant ions \cite{so2024trappedion}.
This paper is organized as follows. In section \ref{sec_model}, we introduce the model for excitation transfer. Section \ref{two_monomer_sec} focuses on the transfer rate of an initially delocalized excitation in a dimer, where each monomer is realized by two ions, both in the perturbative and non-perturbative regimes. Additionally, we investigate the effect of static disorder \crr{and noise} on the transfer rate.
In section \ref{sec_temp}, we investigate the effect of temperature on transfer rates. In section \ref{sec_trimers}, we discuss how this model can be extended to longer chains with more monomers and with monomers composed of more than two qubits. In section \ref{sec_ions}, we propose an experimental implementation of the model in a trapped-ion simulator. Finally, in section \ref{sec_outlook}, we summarize our findings and provide an outlook for future studies.

\section{Model}\label{sec_model}


We consider a one-dimensional system of $N$ spin-$1/2$ qubit ions interleaved with $M$ coolant ions, which are used to perform reservoir engineering on the shared bosonic mode \cite{so2024trappedion}, as shown in Fig.~\ref{Fig1}(a). The qubit ions interact via long-range spin exchange interactions $J_{ij}$ \cite{monroe2021programmable} and are subject to local energy shifts $\epsilon_i$. At the same time, the two-level systems are coupled to the collective bosonic mode defined by creation (annihilation) operators $a^\dagger (a)$ with frequency $\omega$, resulting in the total Hamiltonian:
\begin{eqnarray}
H_{\rm s} &=& H_{\rm el} + H_{\rm v} + H_{\rm el-v} \nonumber\\
H_{\rm s}&=&\sum_i^N \frac{\epsilon_i}{2} \sigma^z_i +  \sum_{i<j}^N J_{ij} (\sigma_i^+ \sigma_j^- + \sigma^-_i\sigma^+_j ) \nonumber \\ 
&+& \omega a^\dagger a +  \sum_{i}^N \frac{g_{i}}{2} \sigma^z_i  (a^\dagger + a).
\label{eq_H_model}
\end{eqnarray}
The Hamiltonian $H_{\rm el}$ encodes the electronic degrees of freedom with the energy landscape that determines the roles of the molecular sites and their mutual electronic couplings. We assume a power-law decay for the electronic couplings $J_{ij}=J/d_{ij}^p$, that can be natively realized in trapped-ion systems \gp{with $J>0$} \cite{monroe2021programmable}. Here, $d_{ij}$ is the distance between the qubit ions, and $p$ is the power-law exponent. Systems in $\mathcal{D}$ dimensions with $p\leq 2\mathcal{D}+1$ have generated a lot of interest recently because deriving tight bounds on the ultimate speed of quantum information transfer is challenging in this regime \cite{Kastner2011, Storch_2015, Guo2020, Tran2021, Jameson_2024}. Here, we consider $p=1$ for most results, which is well within this region\crr{. While the interactions with $p=1$ possess a longer range than the dipole-dipole couplings $(p=3)$, typically assumed in Frenkel exciton models, they can have physical relevance in the case of emitters separated by distances larger than the characteristic photon wavelengths \cite{Daniels2003resonance, Scholes2020, Giovanni2021}. We also comment on how our results extend to the case of $p=3$ in Appendix \ref{app_couplings}.}
The Hamiltonian $H_{\rm v}$ describes a collective vibrational mode of motion that can be encoded in a normal mode of the ion chain, defining a reaction coordinate $y/y_0=1/2(a^\dag+a)$, with $y_0=\sqrt{\hbar/2m\omega}$, and $m$ being the ion mass \cite{Schlawin2021}. Finally, $H_{\rm el-v}$ represents the interaction between electron and vibration with site-dependent couplings $g_i$.

The total Hamiltonian in Eq.~\eqref{eq_H_model} conserves the total number of spin excitations $\Sigma_z=\sum_i \sigma_i^z$. \gp{In this work, we focus on the \emph{single excitation manifold}, where the allowed number of spin states is reduced to $N$}. For example, in the two-monomer case with two sites per monomer (see Fig.~\ref{Fig1}), the two donor states are given by $\vert \! \uparrow \downarrow \downarrow \downarrow \rangle$ and $\vert \! \downarrow \uparrow \downarrow \downarrow \rangle$, while the two acceptor states are given by $\vert  \! \downarrow \downarrow \uparrow \downarrow \rangle$ and $\vert \! \downarrow \downarrow \downarrow \uparrow \rangle$.  

Here, we consider a single, damped bosonic mode to model the relaxation driven by the interaction with the environment. This setting can be realized experimentally in trapped-ion systems through the process of sympathetic cooling  \cite{Schlawin2021, so2024trappedion} (see section \ref{sec_ions} for details). The system dynamics is described by a master equation with Lindbladian super-operators $\mathcal{L}_{c}[\rho]$, where $c$ is a generic jump operator:
\begin{eqnarray}
    &\,&\frac{\partial\rho}{\partial t}=-i[H_{\rm s},\rho] + \gamma (\bar{n}+1)\mathcal{L}_{a}[\rho] + \gamma \bar{n} \mathcal{L}_{a^\dagger}[\rho],
    \label{eq_master}\\
    &\,&\mathcal{L}_{c}[\rho]=
    c\rho c^\dagger - \frac{1}{2}\{c^\dagger c,\rho\}.
    \label{eq_Lind}
\end{eqnarray}
Here, $\rho$ is the total density matrix of the system, $\gamma$ is the motional relaxation rate, and $\bar{n}$ is the phonon population determined by the temperature of the bath $k_B T= \omega/\log(1 + 1/\bar{n})$. This description is motivated by the fact that under the assumption of weak damping ($\gamma\ll \omega$), where the relaxation rate is weaker than both the system's vibrational ($\gamma\ll\omega$) and bath thermal energies ($\gamma\beta\ll 1$, with $\beta=1/k_B T$), the dynamics of the spin and the bosonic observables predicted by Eq.~\eqref{eq_master} are a very good approximation of those of a system in contact with an Ohmic bath described by the spectral density $J(\omega)\sim\omega$, a common choice in electron transfer literature \cite{Garg1985, Lemmer2018, Tamascelli2018}.

\section{Two-Monomer model}\label{two_monomer_sec}

Let us consider a model with only two monomers, which has been used as a minimal model of charge transfer in photosynthetic complexes \cite{Yang2005density,POLYUTOV201221,Tiwari2013, Fujihashi2015, Monahan2015}. In this case, each monomer is composed of two sites and separated by $M=2$ coolant ions (see section \ref{sec_ions2}), as shown in Fig.~\ref{Fig1}. In this case, we can use the following notation:
\begin{eqnarray}
    &\vert 1 \rangle = \vert \! \uparrow \downarrow \downarrow \downarrow \rangle, \quad \vert 2 \rangle = \vert \! \downarrow \uparrow \downarrow \downarrow \rangle, \nonumber \\
    &\vert 3 \rangle = \vert \!  \downarrow \downarrow \uparrow \downarrow \rangle, \quad \vert 4 \rangle = \vert \!  \downarrow \downarrow \downarrow \uparrow \rangle.  
    \label{eq_statelabel}
\end{eqnarray}
Expressed in this basis, $J_{ij}$ is given by:
\begin{equation}
 J_{ij} = \begin{pmatrix}  0&J&\frac{J}{4^p}&\frac{J}{5^p}\\
 J&0&\frac{J}{3^p}&\frac{J}{4^p}\\
 \frac{J}{4^p}&\frac{J}{3^p}&0&J\\
 \frac{J}{5^p}&\frac{J}{4^p}&J&0
 \end{pmatrix}.
 \label{eq_Jij}
\end{equation}
We note that apart from their dissipative role, the number of coolants also determines the distance between neighboring monomers, in this case, $d_{23}=3$. 
We refer to the nearest-neighbor terms $J_{12}$=$J_{34}=J$ as the intra-monomer couplings since they connect two donor (acceptor) states. All other $J_{ij}$ terms correspond to inter-monomer couplings. 
In the configuration shown in Fig.~\ref{Fig1}, the dominant inter-monomer coupling is $J_{23} = {J}/{3^{p}}$. The choice of two coolant ions separating monomers is motivated by the following reasons: \emph{(i)} it imposes a hierarchy in the electronic configuration where the inter-monomer couplings are smaller than the intra-monomer ones \cite{Mattioni2021} without making the former too small; \emph{(ii)} it is more convenient to have even chains in the experimental implementation proposed here (see section \ref{sec_ions}). We keep coolant configuration when considering longer chains in section \ref{sec_trimers}. In the following, we study the transfer of a single excitation from donor sites to acceptor sites as a function of different Hamiltonian parameters.

For simplicity, we consider the symmetric case $\epsilon_1 = \epsilon_2 = -\epsilon_3 = -\epsilon_4 = \epsilon$. With this choice, both donor and acceptor states are shifted symmetrically with respect to zero energy, and the energy gap between donor and acceptor sites is $\epsilon$.
This configuration facilitates the irreversible transfer into the acceptor sites in the dissipative dynamics. Unless specified otherwise, we consider $p=1$ in the following. 
The main goal of this analysis is to understand whether and, if so, which entangled delocalized initial donor states enable faster transfer compared to the initial product states.

\begin{figure*}[t!]
\includegraphics[width=0.95\textwidth]{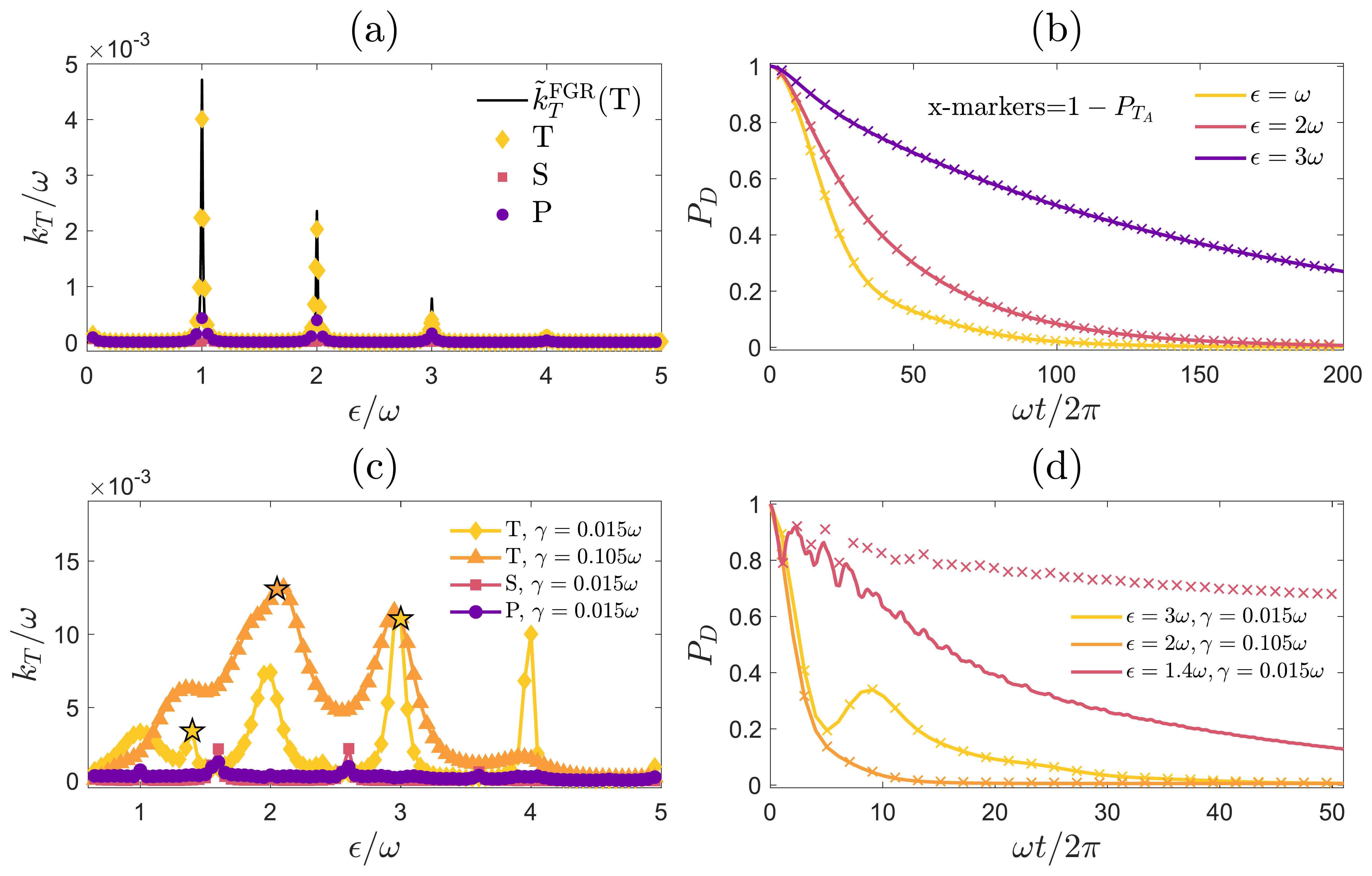}
\caption{ {\bf Excitation transfer: }Transfer rate $k_T$ in the perturbative ((a) and (b)) and non-perturbative regimes ((c) and (d)). 
{\bf (a)} Transfer rate $k_T$ as a function of $\epsilon$ in the perturbative regime ($J,\gamma \ll \omega$) for the different initial states $\vert T_D \rangle$ (T), $\vert S_A \rangle$ (S), and $\vert 2 \rangle$ (P). The black solid line represents the transfer predicted for initial state $\vert T_D \rangle$ using the Fermi Golden Rule $k_T^{\rm FGR}$. Note that we defined $\tilde{k}_{T}^{\rm FGR}=\frac{1}{5}k_{T}^{\rm FGR}$ where we rescale by a factor of $\frac{1}{5}$ to display it clearly in the same plot. 
{\bf(b)} Dynamics at the highest resonances of (a). Solid lines represent the population in the donor sites $P_D$, while the x-markers illustrate the behavior of $1-P_{T_A}$ with $P_{T_A}$ being the population of $\vert T_A \rangle$. All curves correspond to the initial state $\vert T_D \rangle$.  The parameters for (a) and (b) not specified in the plots are $g=\omega$, $J = 0.03 \omega$, $\gamma=0.015 \omega$, $p=1$, and $\bar{n}=0.01$.
{\bf(c)} Transfer rate $k_T$ as a function of $\epsilon$ in the non-perturbative regime $(J\sim \lambda)$ for different initial conditions and parameter values. The parameter sets of data points marked by stars are analyzed further in panel (d). 
{\bf (d)} Dynamics of the donor population for the parameters signaled by stars in (c). Solid lines and x-markers are defined as in (b). The parameters for (c) and (d) not specified in the plots are $g=\omega$, $J = 0.3 \omega$, $p=1$, and $\bar{n}=0.01$.}
\label{Fig2}
\end{figure*}

\subsection{Perturbative regime}\label{sec_pert}


We first consider the perturbative regime in which the electronic coupling is weak compared to the other energy scales in the model ($J\ll g_i,\epsilon,\omega$). In this case, the total Hamiltonian in Eq.~\eqref{eq_H_model} can be divided in $H_{\rm tot} = H_0 + H_J$, where the electronic coupling defined by Eq.~\eqref{eq_Jij} is the perturbation, and the unperturbed Hamiltonian is given by:
\begin{eqnarray}
H_0 &=& \frac{\epsilon}{2} (\vert 1 \rangle \langle 1 \vert + \vert 2 \rangle \langle 2 \vert - \vert 3 \rangle \langle 3 \vert - \vert 4 \rangle \langle 4 \vert) \nonumber \\
&+ & \omega a^{\dagger}a + \sum_{j=1}^4 \frac{g_j}{2} (a+a^{\dagger})\vert j \rangle \langle j \vert,
\label{H01}
\end{eqnarray}
The unperturbed Hamiltonian in Eq.~\eqref{H01} is diagonal in the basis $\{ \vert j \rangle \}$, and thus the unperturbed eigenstates are determined by the diagonalization of the bosonic part of the Hamiltonian: 
\begin{equation}
    H_{B,j} = \omega a^{\dagger}a + \frac{g_j}{2} (a+a^{\dagger}).
    \label{eq_bosonic_Ham}
\end{equation}
As shown in Appendix \ref{app_A}, this Hamiltonian corresponds to a displaced harmonic oscillator with displacement $\alpha_j=-g_j/2\omega$ (see Fig.~\ref{Fig1}(b)). The eigenstates of the bare system are $\vert \psi_{j,n} \rangle =\vert j \rangle \otimes \vert n,\alpha_j \rangle$, where $\vert n, \alpha_j \rangle$ denote the Fock states of the displaced operator at site $j$, and the states $\ket{j}$ are defined in Eq.~\eqref{eq_statelabel}. The eigenenergies of $H_0$ are therefore given by:
\begin{equation}
    E_{j,n} = n \omega - \frac{g_j^2}{4 \omega} + \frac{\epsilon_j}{2}
    \label{eq_E_0},
\end{equation}
where $n$ is the number of bosonic excitations. It is convenient to define the reorganization energy for $j$-th site $\lambda_j=g_j^2/\omega$, which quantifies the amount of energy needed to displace the motional wavepacket by $g_j/\omega$ without any excitation transfer.

Since $H_J$ is a perturbation, we can express the transfer rate between the donor and acceptor states using the Fermi golden rule (FGR) \cite{Garg1985, Schlawin2021}. Considering an initial state $\vert \psi_{i,0} \rangle$ with zero bosonic excitations, the transfer rate to a final state $\vert \psi_{f,n} \rangle$ (see Appendix \ref{app_A}) can be expressed in the zero temperature case as:
\begin{equation}
 k_T^{if}/\gamma = \sum_{n=0}^{\infty}\frac{\vert J_{if} \vert^2}{(\Delta E_{if})^2 + \gamma^2/4}\left(\frac{\exp(-\tilde{n}) \tilde{n}^{n}}{n!}\right),
 \label{eq_FGR}
\end{equation}
where the Franck-Condon factor $|\langle{n,\alpha_f}\ket{0,\alpha_i}|^2$ follows a Poissonian distribution determined by $\tilde{n} = (\alpha_f - \alpha_i)^2 = (g_f/2\omega - g_i/2\omega)^2$. Here, the dissipation is taken into account by considering a Lorentzian broadening of the delta functions in the original FGR \cite{Garg1985, Schlawin2021}. The total transfer rate to the acceptor site is defined by the sum over all possible acceptor states $\vert \psi_{f,n} \rangle$ as $k_T^{\rm FGR}=\sum_{f \in A} k_{T}^{if}$. In this regime, the transfer is maximized when the energy is conserved, namely when $\Delta E_{if} = E_{f,n} - E_{i,0}=0$, which leads to the resonance condition $\epsilon=n\omega + (g_f^2/4\omega - g_i^2/4\omega)$. 




In order to gain an intuitive understanding of the role of delocalized states in the transfer process, it is instructive to take the following choice of spin-boson coupling $g_1=g_2=-g_3=-g_4=g$. \crr{Although a seemingly more natural choice would have been to choose all couplings $g_j$ to also share the same sign, such parameter sets would, however, severely lower the transfer rates for any $\epsilon > 0$, as shown in Eq.~\eqref{eq_FGR}. This parameter choice can be implemented in a trapped-ion setup as described in Section~\ref{sec_ions}}.  In this special parameter choice, we can include the intra-monomer nearest neighbor coupling terms in the unperturbed Hamiltonian, that is:
\begin{eqnarray}
H_0 &=& \omega a^{\dagger}a + \frac{\epsilon}{2} (\vert 1 \rangle \langle 1 \vert + \vert 2 \rangle \langle 2 \vert - \vert 3 \rangle \langle 3 \vert - \vert 4 \rangle \langle 4 \vert) \nonumber \\
&&+ \frac{g}{2} (a+a^{\dagger})(\vert 1 \rangle \langle 1 \vert + \vert 2 \rangle \langle 2 \vert - \vert 3 \rangle \langle 3 \vert - \vert 4 \rangle \langle 4 \vert)  \nonumber \\
&&+ J(\vert 1 \rangle \langle 2 \vert + \vert 2 \rangle \langle 1 \vert + \vert 3 \rangle \langle 4 \vert + \vert 4 \rangle \langle 3 \vert).
\label{H0J}
\end{eqnarray}
The unperturbed vibrational eigenstates are unchanged, but the spin eigenstates are now defined by the triplet and singlet excitonic states on the donor and acceptor sites, respectively:
\begin{eqnarray}
    &&\vert T_D \rangle = \frac{1}{\sqrt{2}}(\vert 1 \rangle + \vert 2 \rangle), \vert S_D \rangle = \frac{1}{\sqrt{2}}(\vert 1 \rangle - \vert 2 \rangle ),\;\;\;\; \nonumber \\
    &&\vert T_A \rangle = \frac{1}{\sqrt{2}}(\vert 3 \rangle + \vert 4 \rangle), \vert S_A \rangle = \frac{1}{\sqrt{2}}(\vert 3 \rangle - \vert 4 \rangle ).\;\;\;\;
\end{eqnarray}
The new eigenstates can still be defined as $\vert \psi_{j,n}\rangle$ with $j=T_D,S_D,T_A,S_A$. In this excitonic basis, the FGR in Eq.~\eqref{eq_FGR} considers indices $i=T_D, S_D$ and $f=T_A, S_A$. These eigenstates have energies identical to those on Eq.\eqref{eq_E_0} with the difference that the triplet (singlet) states are shifted by $+J\;(-J)$. \crr{A pictorial representation of the energetics in this new basis is provided in Fig.~\ref{Fig1}(b).} In this new notation, the perturbation Hamiltonian $H_J$ now corresponds to all the inter-monomer coupling terms \crr{(see green arrows in Fig.~\ref{Fig1}(b))}:
\begin{eqnarray}
    H_J &=& J_{TT} (\vert T_D \rangle \langle T_A\vert + \textnormal{h.c.}) \nonumber \\
    &\quad& +\;J_{ST} (\vert S_D \rangle \langle T_A\vert + \textnormal{h.c.}) \nonumber \\
   &\quad& +\;J_{TS} (\vert T_D \rangle \langle S_A\vert + \textnormal{h.c.}) \nonumber \\
   &\quad& +\;J_{SS} (\vert S_D \rangle \langle S_A\vert + \textnormal{h.c.}).
    \label{HperTS}
\end{eqnarray}
\crr{We note that there are no intra-monomer couplings in this basis, thanks to the values chosen for $g_j$}. The coupling strengths are explicitly given by:
\begin{eqnarray} \label{couplings_ts}
    &&J_{TT} = \left(\frac{1}{2}\frac{J}{3^{p}} + \frac{J}{4^p} + \frac{1}{2}\frac{J}{5^{p}} \right), \nonumber \\
    &&J_{TS}= -J_{ST} = \left(\frac{1}{2}\frac{J}{3^{p}} -\frac{1}{2}\frac{J}{5^{p}} \right), \nonumber \\
    &&J_{SS} = \left(-\frac{1}{2}\frac{J}{3^{p}}+\frac{J}{4^{p}} -\frac{1}{2}\frac{J}{5^{p}} \right).
\end{eqnarray}
A plot of the ratio of these inter-monomer couplings as a function of $p$ is shown in Fig.~\ref{Fig1supp} in Appendix \ref{app_couplings}. For values of $p \leq 3$, the coefficient $J_{TT}$ is the dominant term, with its value being significantly larger than both $J_{ST}$ and $J_{SS}$. This means that if the system is initialized in a triplet donor state $\vert \psi_{T_D,0} \rangle$, the triplet-to-triplet transfer will be faster than the one associated with any initial donor product state since $J_{TT}$ is greater than any individual inter-monomer coupling. Additionally, because both triplet states, $\vert T_D \rangle$ and $\vert T_A\rangle$, have their energy shifted by $+J$, the resonance condition remains $\epsilon=n\omega$, which is independent of the spin-phonon coupling strengths $g_j$ for the current parameter choice \crr{(see Fig.~1(b))}. 

We check this prediction numerically by integrating Eq.~\eqref{eq_master} for different initial states up until $\omega t_{\rm sim}/2 \pi = 100$ \crr{(see Appendix \ref{app_simulations} for details on the dynamics of the bosonic degrees of freedom and truncation of the bosonic Hilbert space)}. In Fig.~\ref{Fig2}(a-b), we show the transfer rate $k_T$ from donor to acceptor as a function of $\epsilon$ for $J = 0.03\omega$, which is well within the perturbative regime. The transfer rate obtained numerically is defined in terms of the population in the donor state $P_{D} = \langle (\vert 1 \rangle \langle 1 \vert + \vert 2 \rangle \langle 2 \vert)\rangle$  as \cite{Skourtis_1992, Schlawin2021, so2024trappedion}:
\begin{equation}
    k_T = \frac{\int_{t=0}^\infty P_D(t)dt}{\int_{t=0}^\infty t P_D(t)dt}.
    \label{transferrate}
\end{equation}
This definition can characterize the transfer rate for arbitrary donor population dynamics and depends on both the speed at which the system reaches the steady state and the value of the steady-state population itself. In the special case when the donor population dynamics follows an exponential decay, we can also define an equilibration rate $\Gamma$ as:
\begin{equation}
P_D = (1-P_D^{SS}) \exp(-\Gamma t)+ P_D^{SS},
\label{eq_Gamma_def}
\end{equation}
where $P_D^{SS}$ is the donor population in the steady state. Differently from the transfer rate $k_T$, the equilibration rate $\Gamma$ characterizes only how fast the steady state is reached, independent of the specific value of $P_D^{SS}$. For a complete population transfer $(P_D^{SS}=0)$ and exponential decay, the two rates coincide.



Fig.~\ref{Fig2}(a) shows the expected peaks in the transfer rate at the resonances $\epsilon = n \omega$ for triplet, singlet, and product states. The relative height of the peaks is determined by the Franck-Condon factor, which, following the Poissonian distribution dependent on $g$ and $\omega$ (see Eq.~\eqref{eq_FGR}), peaks at $\tilde{n}$=1 since $g=\omega$ for all the simulations. \crr{We note that we choose this value of $g$ to explore the regime in which the donor and acceptor sites are described by displaced potential energy surfaces with quantized vibrational levels (see Fig.~\ref{Fig1}(b)). The weakly coupled regime with $g/\omega \lesssim 0.1$ has been explored before, see, for example, Refs.\cite{Gorman2018,so2025quantumsimulationschargeexciton}}. The numerically obtained transfer rates qualitatively agree with those predicted by the FGR, as shown by the black solid line (note that we display $\tilde{k}_T^{\rm FGR}=k_T^{\rm FGR}/5$ in the plot). However, the FGR predicts a higher transfer rate, which can be attributed to the fact that this expression is derived under the assumption $\gamma \gg J$~\cite{Skourtis_1992}. This condition differs from the parameters used in panel (a), where $J = 0.03 \omega$ and $\gamma = 0.015 \omega$.

Remarkably, in the perturbative regime, the transfer rate of a symmetric spin triplet superposition ($\ket{T_D}$) is significantly higher than the transfer rate starting in either an antisymmetric singlet superposition $\ket{S_D}$ or a product state $\vert 2 \rangle$. This observation can be understood as follows: since the transfer rate is proportional to $J_{ij}^2$ in the perturbative FGR regime, the $\ket{T_D}\rightarrow\ket{T_A}$ transfer is more efficient than the $\ket{S_D}\rightarrow\ket{S_A}$ case by a factor proportional to $(J_{TT}/J_{SS})^2\gg 1$. Similarly, the triplet state will transfer faster than the most favorable product state $\ket{2}$ by at least $(J_{TT}/J_{23})^2> 1$ (see Appendix \ref{app_couplings}).
\gp{We note that the role of triplet and singlet states are determined by the sign of the nearest-neighbor electronic coupling $J>0$ and that they can be exchanged by programming the electronic couplings to have alternating signs ($J_{ij}\rightarrow(-1)^{|i-j|}|J_{ij}|$) \cite{Hestand2018}.}

In Fig.~\ref{Fig2}(b), we show the transfer dynamics from the initial state $\vert \psi_{T_D,0} \rangle$ for the three main resonances $\epsilon = \omega, 2\omega, 3\omega$. $P_D$ shows an approximately exponential decay as expected from Eq.~\eqref{eq_FGR} into the acceptor state. Importantly, Fig.~\ref{Fig2}(b) also shows $1 - P_{T_A}$ (signaled by x-markers), where $P_{T_A} = \langle \vert T_A \rangle \langle T_A \vert \rangle$, confirming that the transfer into the acceptor is a coherent process in which not only the population but the full quantum state is transferred to the acceptor site. \crr{Given that $\ket{T_D}$ and $\ket{T_A}$ are maximally entangled states of the qubits in the donor and acceptor, respectively, the dynamics can also be understood in terms of entanglement transfer, see Appendix \ref{app_entanglement} for more details.} It is worth noting that, when starting in the triplet state, the system behaves approximately as a two-level system defined by the $\{\ket{T_D},\ket{T_A}\}$ subspace. 
In this electronic two-level description, an initial state $\ket{\psi_{T_D,0}}$ is coupled to state $\ket{\psi_{T_A,n}}$ through the electronic coupling $J_{TT}$ when $\epsilon = n\omega$. Decay through the relaxation rate $\gamma$ couples the latter to lower energy vibrational states $\ket{\psi_{T_A,m}}$ with $m<n$. Eventually, the system decays to state $\ket{\psi_{T_A,0}}$ where, in the case of near-zero temperature, it gets trapped as this state is off-resonant with respect to any donor state (see section ~\ref{sec_temp} for the effects of finite temperature). For the population to remain trapped in $\vert \psi_{T_A,0} \rangle$, it is necessary that $\epsilon > 2J$, otherwise, the state $\vert \psi_{S_D,0} \rangle$ could get populated even if this process is slower (rate proportional to $J_{ST}^2$). We adopt the condition $\epsilon > 2J$ throughout this work. 

For all these reasons, the results in the non-perturbative regime presented in Fig. \ref{Fig2}(a) show good agreement with previous studies on electron-transfer models for a single donor and acceptor site \cite{Schlawin2021,so2024trappedion}. The more complex structure of the monomer, however, plays an important role when the system is outside the perturbative regime, which will be addressed in the next section.

\subsection{Non-Perturbative regime}\label{sec_non_pert}

When the intermonomer coupling becomes comparable to the reorganization energy $J\sim\lambda$, the inter-monomer coupling cannot be treated as a perturbation. This non-perturbative regime can be accessed by increasing $J$ or decreasing the power-law exponent $p$. In Fig.~\ref{Fig2}(c), we consider the first case as we take $J=0.3\omega$ while keeping $p$ fixed. The role of $p$ will be discussed in more detail in the following sections and in Appendix \ref{app_couplings}. 


In this regime, we note that the resonances at $\epsilon= n \omega$ are still present despite the relative height not following the pattern expected from the effective vibronic couplings determined by the Franck-Condon factors. Additionally, new resonances appear at non-integer values of $\omega$. These resonances can be understood as a transfer from an initial state $\vert \psi_{T_D,0} \rangle$ to a final state $\vert \psi_{S_A}, n \rangle$. The energy difference between such states is given by $\Delta E_{if} = n \omega - 2J$, which for the value of $J = 0.3 \omega$ considered here, yield resonances at $\epsilon/\omega= 0.4, 1.4, 2.4$, and so on. This is confirmed numerically in Fig.~\ref{Fig2}(c) (see yellow diamonds). Similarly, if the system is initialized in state $\ket{\psi_{S_D,0}}$, the transfer to the state $\ket{\psi_{T_A,n}}$ becomes resonant at values $\epsilon/\omega = 0.6, 1.6, 2.6, \dots$  (pink squares). Although these processes are still slower than the triplet-to-triplet transfer, the height difference between the triplet-triplet and the singlet-triplet peaks is smaller than the ratio $J_{TT}^2/J_{TS}^2$, illustrating the departure from the perturbative description of the dynamics. To understand this behavior in-depth, we show the time dynamics of the transfer for different values of $\epsilon$ in Fig.~\ref{Fig2} (d). In the case of $\epsilon = 1.4 \omega$ (pink lines and markers), the total donor population is not predicted only by the population in the triplet acceptor ($P_D \neq 1-P_{T_A}$), which means that the population is transferred to both the triplet and singlet acceptor states. This is the result of the competition between two processes: first, the transfer from $\vert \psi_{T_D,0}\rangle$ to $\vert \psi_{T_A,n} \rangle$ features a larger coupling ($\propto J_{TT}$) but the energy difference between them is finite since $\epsilon$ is not an integer of $\omega$ and acts as an effective detuning; second, the transfer from $\vert \psi_{T_D,0} \rangle$ to $\vert \psi_{S_A,n}\rangle$ has a lower coupling ($\propto J_{TS}$) but the states are resonant. The combination of these two processes results in an overall higher transfer rate $k_T$ compared to the one expected in the perturbative regime.

Similarly, the triplet-to-triplet transfer is favored for values of $\epsilon$ being integers of $\omega$ both from a large transfer rate and a resonant energy, making the transfer at these parameter values considerably faster than any other configuration with initial singlet or product states. As shown in Fig.~\ref{Fig2} (d), at these resonances, the triplet state gets fully transferred to the acceptor site as signaled by the x-markers ($P_D \simeq 1-P_{T_A}$). 

In Eq.~\eqref{eq_FGR}, we incorporated the motional relaxation rate $\gamma$ as an effective broadening of the transfer resonances. In the non-perturbative regime, the effect of $\gamma$ is similar as shown numerically in Fig.~\ref{Fig2}(c) with orange triangles. We note that all resonances broaden as $\gamma$ increases, with the peaks slightly shifting from the expected perturbative values. Comparing the results for low and high $\gamma$ (triangles and diamonds), we note that for certain resonances, an increase of $\gamma$ enhances the transfer rate ($\epsilon = 2\omega$), while for other cases, it has the opposite effect in the $\epsilon = 4 \omega$ case (see also Fig.~\ref{Fig2}(d), yellow line and markers). This behavior poses the question of how we can optimize the transfer rate $k_T$ by modifying the different parameter values, which will be discussed next.

 \begin{figure*}[t!]
    \centering
    \includegraphics[width=\linewidth]{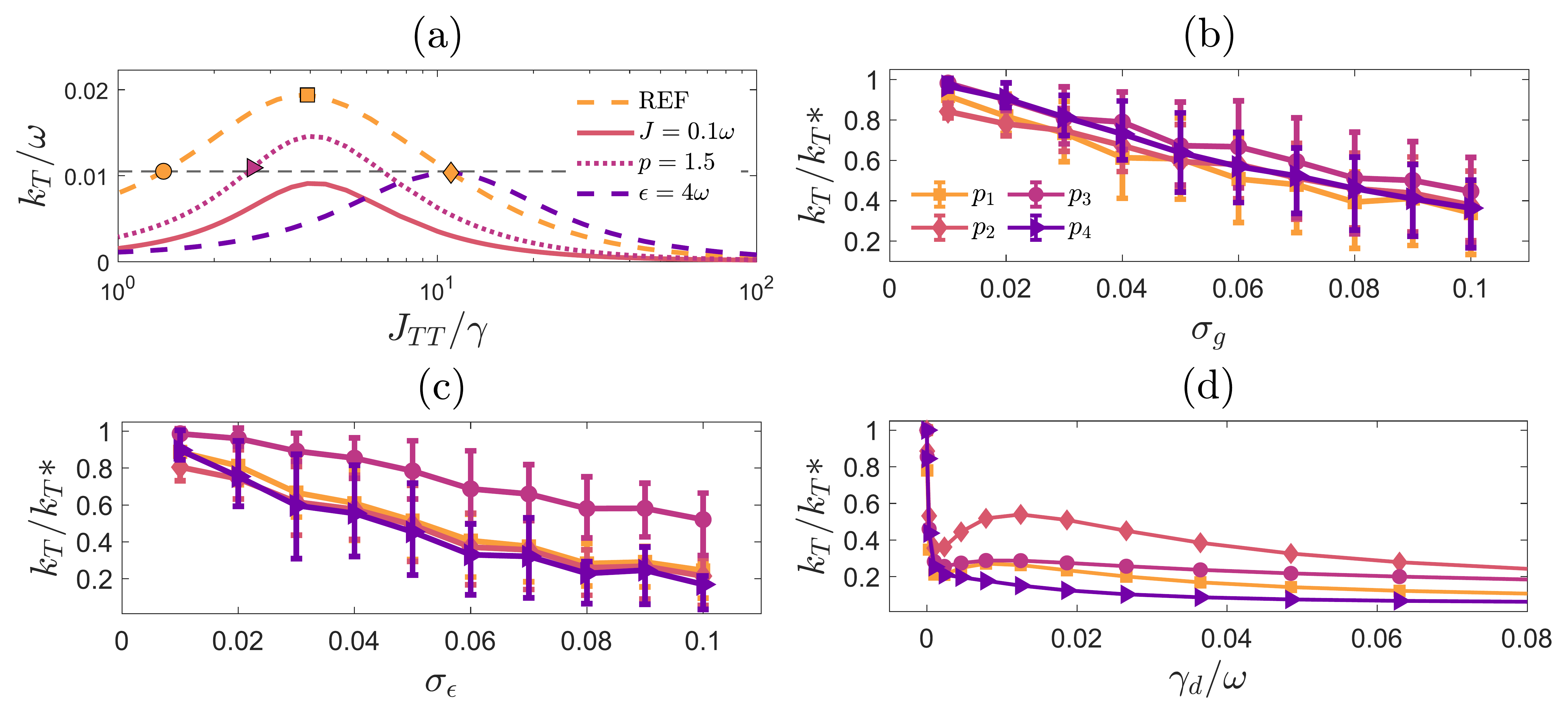}
    \caption{\textbf{Optimal transfer and static disorder}: 
    {\bf (a)} The transfer rate as a function of $J_{TT}/\gamma$ is presented for different parameter conditions. $J$ is kept fixed for each plot and $\gamma$ is varied. The reference parameter set (REF in the legend) corresponds to $J=0.3\omega$, $p=1$, $g=\omega$, $\epsilon=3\omega$, and $\bar{n}=0.01$. For all other curves, one of these parameters is modified, as indicated in the legend, while all others remain the same. Four $\gamma$ values are chosen for further analysis (see Table~\ref{table:parameter sets} for the exact values of $\gamma$). These parameter sets are labeled as $p_1$, $p_2$, $p_3$, and $p_4$, corresponding to the square, circle, diamond, and triangle markers, respectively. The horizontal dashed line is used to illustrate that $p_2$, $p_3$, and $p_4$ yield almost identical transfer rates.
    {\bf (b)} Transfer rate as a function of the standard deviation of the static disorder in the $g_j$ couplings. $k_T*$ represents the transfer value without disorder. Each data point is the average over one hundred disorder realizations, and error bars represent the 25th and 75th percentiles. 
    {\bf (c)} Same as in (b) but for disordered on-site energies $\epsilon_j$.
    \crr{{\bf (d)} Transfer rate as a function of the dephasing rate $\gamma_d$. $k_T*$ represents the transfer rate value without noise. All markers and colors are defined as in (b) and (c).}}
    \label{Fig3}
\end{figure*}

\subsection{Optimal Transfer, Static Disorder\crr{, and Noise}}


As shown in Fig.~\ref{Fig2}(c) and (d), the ratio between the electronic coupling $J_{TT}$ and the relaxation rate $\gamma$ is a crucial parameter in determining the transfer rate in the non-perturbative regime. Simply increasing the electronic coupling $J_{TT}$ might induce oscillations between donor and acceptor before the steady state is reached, therefore reducing the transfer rate. On the other hand, setting a larger relaxation rate $\gamma$ broadens the vibrational levels making the resonant process less prominent. Consequently, all else being equal, we expect an optimal transfer when $J_{TT}$ and $\gamma$ are of the same order, similar to a critically damped classical oscillator. This is reminiscent of what is believed to happen in light-harvesting materials, where the relaxation of the environment occurs on similar timescales as the coherent electronic couplings \cite{Fassioli2014}.

In Fig.~\ref{Fig3}(a), we report numerical results of the transfer rate $k_T$ as a function of $J_{TT}/\gamma$. Each transfer rate curve is obtained by varying $\gamma$ while keeping all other parameters fixed. We note that, given $g$ and $\epsilon$, the transfer peaks at the same value of $J_{TT}/\gamma$ regardless of the specific values of $J$, $\gamma$, and $p$. The position of the peak, however, can be shifted if a different value of $\epsilon$ or $g$ is chosen (see purple dashed line). We note that a similar optimization behavior of the transfer was reported when considering a single donor and acceptor states \cite{Schlawin2021,so2024trappedion}.



So far, we considered a symmetric configuration with $|\epsilon_j|=\epsilon$ and $|g_j|=g$, finding that it favors the transfer from triplet to triplet states. However, it is important to consider the effect of static disorder that can impact the transfer rate by localizing the excitation either due to heterogeneity of site energies $\epsilon_j$ or inducing ``dynamical localization'' via the spin-phonon couplings $g_j$ \cite{Fassioli2014}. In both cases, we expect the excitons to localize and the transfer to be suppressed. 

To study the effect of static disorder on the transfer rate, we consider four parameter configurations $p_1$, $p_2$, $p_3$, and $p_4$, signaled with markers in Fig.~\ref{Fig3}(a), \gp{and we average over 100 disorder realizations}. $p_1$ represents the optimal $J_{TT}/\gamma$ ratio, while $p_2$, $p_3$, and $p_4$ are chosen such that they all yield almost identical transfer rates despite representing different parameter values (see horizontal dashed line in Fig.~\ref{Fig3}(a)). 
To model the static disorder, we consider $g_j = \bar{g}_j(1+\delta^g_j)$, with $\bar{g}_1 = \bar{g}_2 = - \bar{g}_3 = -\bar{g}_4 = g$, and $\delta^g_j$ being a Gaussian random variable with zero average and standard deviation $\sigma_g$. 
In Fig.~\ref{Fig3}(b), results for different values of $\sigma_g$ are shown. The vertical axis is normalized with respect to the optimal value without disorder $k_T^*$ for each parameter set $p_j$. As expected, when the disorder increases, the transfer rate is lowered. A site-dependent $g_j$ changes the energy of the vibronic states (see Eq.~\eqref{eq_E_0}), affecting the resonance condition. Furthermore, the disorder in the displacements $g_j/\omega$ associated with the donor (acceptor) sites also reduces the transfer rate by decreasing the Franck-Condon factor. The decrease in transfer rate can be understood by noting that the reorganization energies of each qubit $\lambda_j=g_j^2/\omega$ are, on average, always larger than the electronic coupling $J$ for our parameter choice, resulting in dynamical localization of the excitonic states.

As $\sigma_g$ increases, we note a higher variance of the possible transfer rates.
Indeed, in the presence of static disorder, the different $g_j$ can favor transfer processes that are not resonant when all couplings have the same magnitude $\vert g_j \vert = g$. This may result in the prominence of slower transfer triplet-to-singlet or singlet-to-singlet processes. Moreover, certain disorder realizations can lead to a larger donor population in the steady state, decreasing the transfer rate (see discussion after Eq.~\ref{transferrate} and Appendix \ref{app_disorder}). 
Additionally, the transfer rates in the four configurations in Fig.~\ref{Fig3}(b) are all affected in a similar way by disorder in the spin-phonon coupling $g_j$. However, the optimal parameter set $p_1$, on average, always yields a higher transfer rate in absolute terms since the vertical axis in Fig.~\ref{Fig3}(b) is normalized with respect to $k_T^*$. 

In Fig.~\ref{Fig3}(c), we consider the case of static disorder in the on-site energies $\epsilon_j$, and we assume an identical disorder with standard deviation $\sigma_\epsilon$ for each qubit.
\begin{figure*}[t!]
    \centering
    \includegraphics[width=\linewidth]{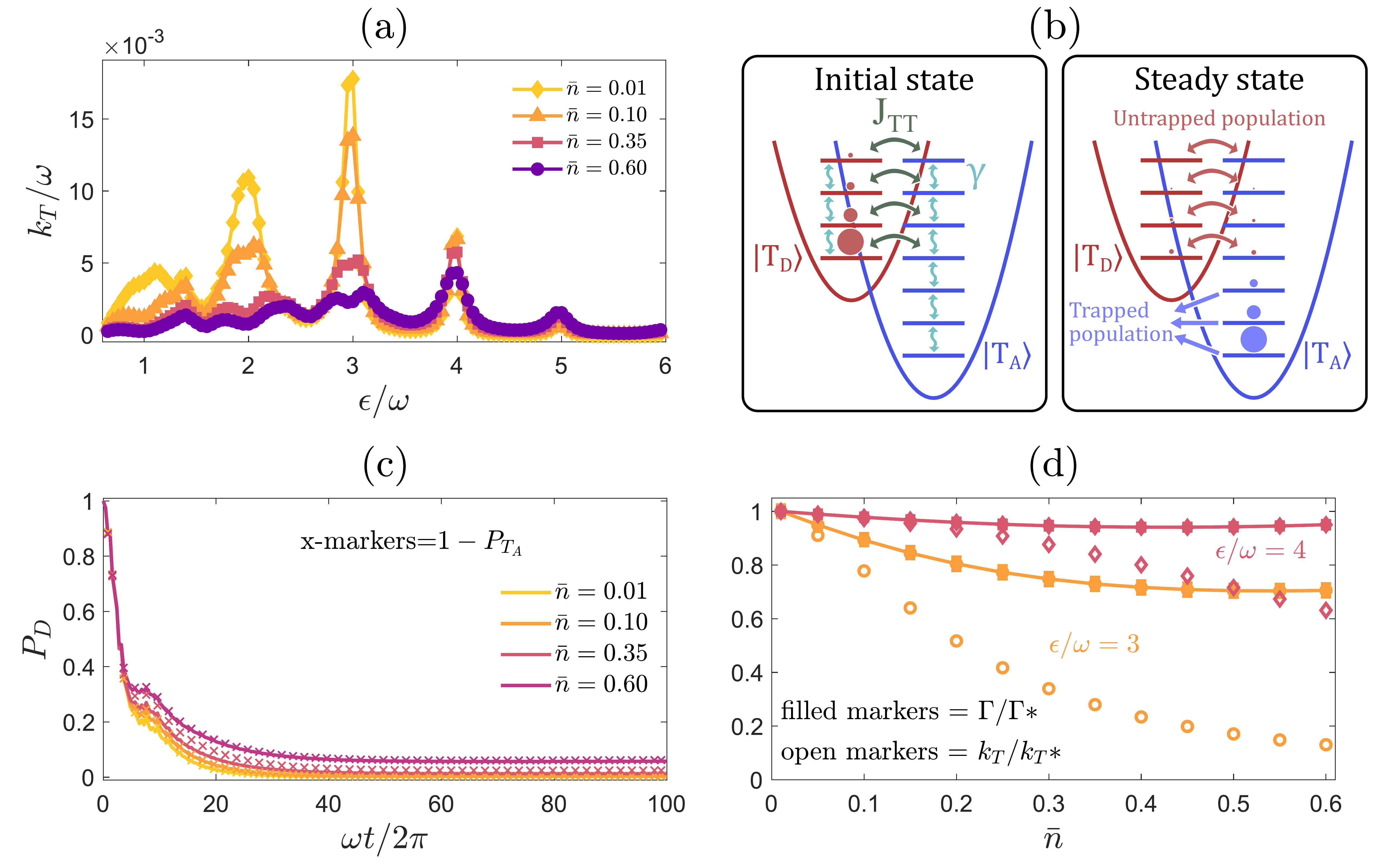}
    \caption{\textbf{Effects of temperature}: 
    {\bf(a)} Transfer rate $k_T$ as a function of $\epsilon$ in the for different values of $\bar{n}$. All other parameters are defined as in parameter set $p_1$ (see Table ~\ref{table:parameter sets}) for this and subsequent panels. 
    {\bf(b)} Schematics of the triplet-triplet transfer for a thermal state. An initial thermal state with electronic state $\vert T_D\rangle$ gets coupled to the acceptor states through the coherent coupling $J_{TT}$. Coherent evolution and dissipative dynamics ($\gamma$) produce a thermal distribution in the acceptor site in the steady state. A portion of the population remains trapped in the acceptor, while the rest remains coherently coupled to some donor states. The ratio of these two populations depends on the value of $\bar{n}$. 
    {\bf(c)} Population of the donor as a function of time for different temperatures. x-markers denote 1-$P_{T_A}$. Here $\epsilon=3\omega$.
    {\bf(d)} Transfer rate $k_T$ and equilibration rate $\Gamma$ as a function of temperature. Open markers denote $k_T/k_T*$ while filled markers denote $\Gamma/\Gamma*$, with $k_T*$ and $\Gamma*$ being the rates at zero temperature. The error bars for filled markers denote the 95\% confidence intervals of the exponential fit. Different colors signal different values of $\epsilon/\omega$.}
    \label{Fig4}
\end{figure*}
A difference in $\epsilon_j$ only modifies the energy shifts of the different states, differently from the site-dependent $g_j$ couplings that also modify the Franck-Condon factors. In general, we expect to still have delocalized states contributing to the dynamics if the energy variations are small compared to the electronic coupling ($\sigma_{\epsilon}\ll J/\omega $), while a more disordered energy landscape ($ \sigma_{\epsilon}\gg J/\omega$) localizes the excitations, decreasing the transfer rate. Additionally, as shown in Fig.~\ref{Fig2}(c), a faster relaxation rate $\gamma$ broadens the width of the resonances in the spectrum, therefore making the resonance conditions more robust against variations in $\epsilon_j$. The results in Fig.~\ref{Fig3}(c) confirm this behavior as the parameter set $p_3$ (circle markers) with $\gamma \approx 0.1 \omega$ exhibits a considerably slower decrease in the transfer rate than all other parameter choices with $0.01  \lesssim \gamma/\omega \lesssim 0.04$. Interestingly, at high disorder $\sigma_\epsilon =0.1$, the parameter set $p_3$ with larger $\gamma$ exhibits a larger transfer $k_T$ also in absolute terms compared to the other configurations, differently from the case of small disorder, when the $p_1$ configuration is optimal. Consequently, for fixed electronic coupling $J_{TT}$, the presence of static disorder changes the relaxation rate $\gamma$ that maximizes the transfer. The interplay of dissipation and disorder has also been explored in the case of Anderson localization \cite{Logan1987} and light-harvesting complexes \cite{Caruso2009}.

\crr{Apart from static disorder, time-dependent noise in the system parameters has shown to be important in different transfer setups \cite{Maier2019, Dwiputra2021, Li_2022Whaley}, especially in the context of environment-assisted quantum transfer (ENAQT), where moderate noise strengths might enhance excitation transfer rates \cite{Mohseni2008, Plenio2008, Caruso2009}. In our case, we incorporate noise by considering the parameters $\epsilon_j(t)$ to change in time. As shown in Ref.~\cite{Chenu2017}, if the fluctuations on $\epsilon_j(t)$ can be described by white-noise stochastic processes, the dynamics can be equivalently modeled by a time-independent Hamiltonian and a set of Lindbladian jump operators of the form $\mathcal{L}_{\sigma_j^z}[\rho]$ (see Eq.~\eqref{eq_Lind}). Here we consider the same approach and add the terms $\gamma_d \sum_{j=1}^N \mathcal{L}_{\sigma_j^z}[\rho]$ to the master equation in Eq.~\eqref{eq_master}, where the dephasing rate $\gamma_d$ is proportional to the width of the white noise. Here, we consider the same dephasing rate for all qubits.

In Fig.~\ref{Fig3}(d) we show the normalized transfer rate as a function of $\gamma_d$ for the four sets of parameters $p_j$ in Table \ref{table:parameter sets}. In general, dephasing leads to a slower transfer rate, as compared to the noise-free case. However, the dependence on $\gamma_d$ is not monotonic for all cases. For parameter sets $p_1$, $p_2$, and $p_3$, the transfer rapidly reaches a minimum, and then it increases again up to a local maximum. Similar cases in which the transfer rate has a non-monotonic behavior as a function of the noise strength have been reported in similar setups \cite{Maier2019, Li_2022Whaley}. We should note, however, that we do not observe a case for the chosen parameter sets, where the presence of noise leads to a faster transfer with respect to the noise-free case. This is expected as the transfer speed in the model studied here is enhanced due to the coherences between the qubits of each monomer, and dephasing naturally destroys these coherences.}



\section{Effect of temperature}\label{sec_temp}

So far, we have considered the bath to be at very low temperatures ($\bar{n}=0.01$) and assumed the system to start from the ground state of the displaced harmonic oscillator, namely, $\vert \psi_{j,0} \rangle$, where $j=T_D$ or $j=S_D$, which is a good approximation for a low-temperature thermal state. In this case, the picture of the population getting fully transferred in the acceptor sites in the steady state is accurate, except for the disordered cases discussed in the previous section. Nonetheless, temperature can play a key role in the transfer of excitons when thermal energies become comparable to the other energy scales of the system \cite{Campaioli_2021}.

Here, we consider the case of higher temperatures, which affect both the steady-state population and the transfer timescales.
We consider thermal energies of the order of the vibrational energy, that is, $0 < k_B T \lesssim \omega$, which translates to $0 < \bar{n} \lesssim 0.7$. We assume the initial state to be a thermal state of the form:
\begin{equation}
\rho_j = \sum_{n} \frac{\bar{n}^n}{(1+\bar{n})^{n+1}} \vert \psi_{j,n} \rangle \langle \psi_{j,n}\vert,
\label{eq_thermalinitial}
\end{equation}
where $\bar{n}$ corresponds to the temperature set by the bath\crr{, and all the calculations are performed for the triplet donor state with $\ket{\psi_{j,n}}=\ket{\psi_{T_{D},n}}$}. We report in Fig.~\ref{Fig4}(a) the transfer rate as a function of $\epsilon$ for different temperatures. First, when the temperature increases, the transfer rate at the resonant peaks (integer values of $\epsilon/\omega$) decreases. However, it is important to point out that the transfer rates at higher values of $\epsilon$ are more resilient to the temperature increase than those for lower $\epsilon$.

We can understand this behavior through the schematics presented in Fig.~\ref{Fig4}(b). Initially, a thermal state (Eq.~\eqref{eq_thermalinitial}) is coherently coupled to a family of states $\{\vert \psi_{T_A,n} \rangle\}$ with coupling strength $J_{TT}$. In the acceptor sites, the dissipative dynamics couples different vibrational states until a thermal distribution is reached in the steady state. Indeed, the heating term in Eq.~\eqref{eq_master} is relevant when $\bar{n}$ is close to 1. 


As shown in Fig.~\ref{Fig4}(b), depending on the energy offset between donor and acceptor, the population in the low-lying acceptor states is not coupled with the states in the donor sites because of the energy penalty. On the other hand, higher thermally populated vibronic acceptor states are resonant with the donor states, resulting in a finite donor population $P_D$ in the steady state. What portion of the population remains in the donor sites in the steady state depends on the bath thermal distribution, determined by $\bar{n}$, and by the value of $\epsilon$. Higher values of donor-acceptor imbalance $\epsilon$ always lead to transfers that are more robust with respect to the temperature because, if all other parameters are fixed, more energy is needed to overcome the energy barrier.


As briefly discussed in section \ref{sec_non_pert}, the transfer rate $k_T$ defined in Eq.~\eqref{transferrate} may differ from the equilibration rate $\Gamma$ defined in Eq.~\eqref{eq_Gamma_def}. The transfer rate $k_T$ can yield a lower value if i) the speed at which the steady state is reached decreases or ii) if the steady-state population of the donor sites $P_D^{SS}$ increases. 
We show in Fig.~\ref{Fig4}(c) the time evolution of the donor population $P_D(t)$ for different temperatures to distinguish between these two causes of the decreased $k_T$ as a function of $\bar{n}$ observed in Fig.~\ref{Fig4}(a). We note that, as expected from the schematics in Fig.~\ref{Fig4}(b), for a fixed value of $\epsilon = 3\omega$, $P_D^{SS}$ increases as $\bar{n}$ increases. However, the time it takes to effectively reach the steady state, quantified by $\Gamma$, is mildly affected by the increase in temperature. 

To highlight this point, we fit the curves on Fig.~\ref{Fig4}(c) to Eq.~\eqref{eq_Gamma_def}. The results of this fit are reported in Fig.~\ref{Fig4}(d), where they are contrasted with the behavior of $k_T$. The response of $\Gamma$ to variations in $\bar{n}$ shows that the donor population equilibration rate is quite robust to temperature in the non-perturbative case. 

This is especially significant in processes whose rates are more relevant than the specific values of $P_D^{SS}$, such as those involving a single elementary step of a more complicated charge transfer reaction. When the donor and/or acceptor are intermediate products of a multi-step process, the population transfer is generally not complete, but the equilibration rate is a crucial factor in determining the total rate of the reaction \cite{scheer2017molecular}. Alternatively, when the donor and acceptor are coupled to large source and drain reservoirs, the equilibration rate is the main factor that defines the dynamics of the entire process \cite{Zhuravel2020}.


\section{Higher complexity monomer structures}\label{sec_trimers}
The model studied in the previous sections represents a simple design compared to models describing excitation transfer in biological and chemical systems, where each monomer is composed of several molecular sites, and multiple monomers participate in the excitation transfer. For example, Fenna–Matthews–Olson and the light-harvesting complex II are trimers, where each monomer has 8 \cite{Levi2015} and 12 molecules \cite{Novoderezhkin2010,
Arsenault2020}, respectively.
In this section, we consider two different ways to add complexity to the model: first, we consider monomers with a larger number of two-level systems per monomer (see inset in Fig.\ref{Fig5}(a)); secondly, we consider a larger number of monomers, each one composed of two qubits (inset in Fig.\ref{Fig5} (c)). 

\subsection{Larger Monomers}

We start with the case where each monomer is composed of three qubits. Here, we denote the number of qubits per monomer by $L$ (in this case, $L=3$). The dimension of the electronic Hilbert space is $N=2L=6$, and it can be spanned in terms of the product states with five spins down and one spin up $\vert j \rangle$, where $j=1,2,3\equiv j_D$ for the donor site and $j=4,5,6\equiv j_A$ for the acceptor site. We adopt a similar parameter choice as in the previous sections with $\epsilon_{j_D}=-\epsilon_{j_A}=\epsilon$ and $g_{j_D}=-g_{j_A}=g$, and we consider the same power-law decay form for $J_{ij}$. Note that the latter implies that now the intra-monomer couplings can have different strengths, for example, $J_{12}>J_{13}$.
\begin{figure*}[t!]
    \centering
    \includegraphics[width=\linewidth]{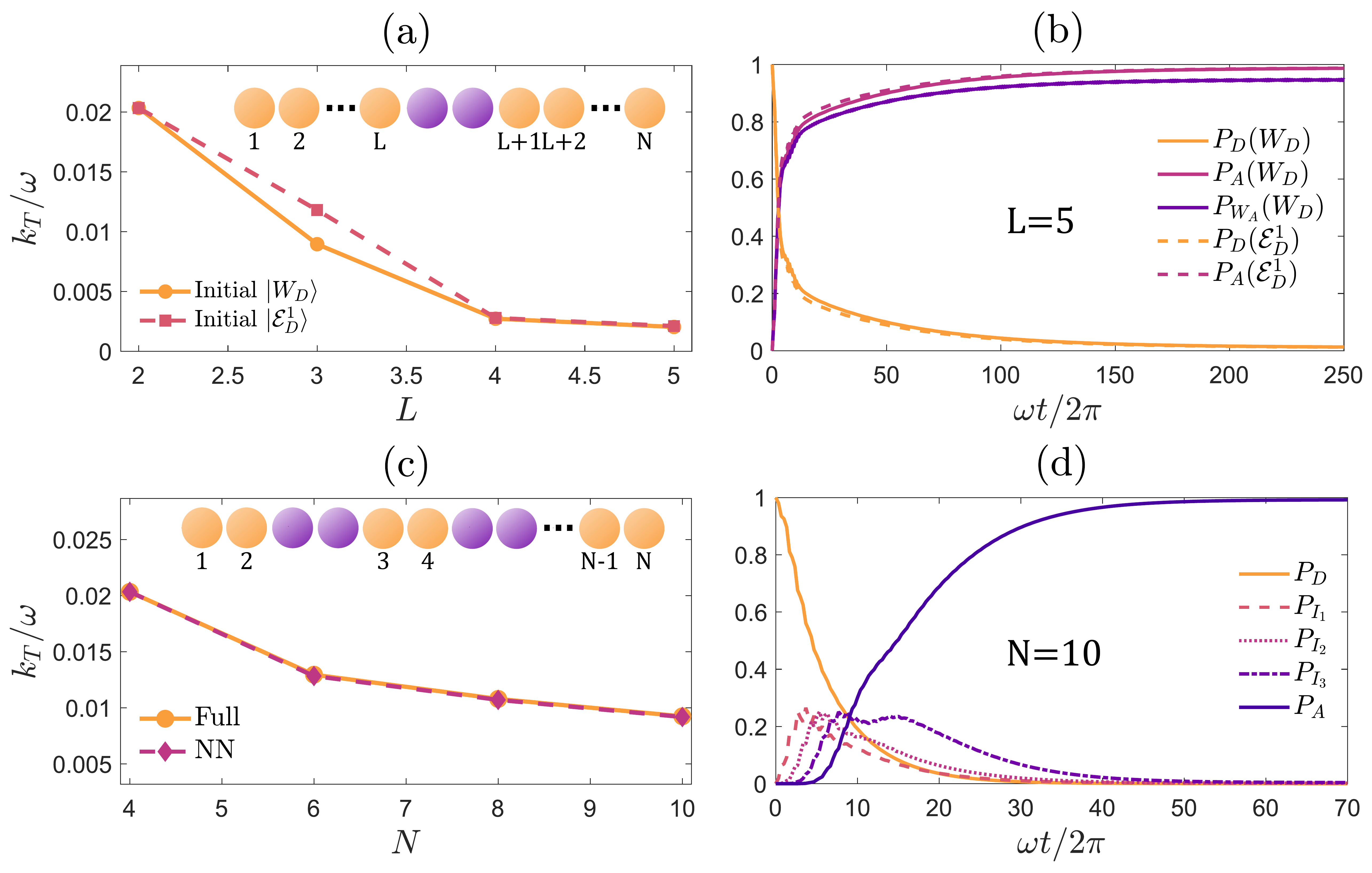}
    \caption{\textbf{Higher complexity monomer structures}: {\bf(a)} Transfer rate as a function of the number of qubits per monomer ($L$). The inset shows a schematic representation of how more two-level systems are incorporated in each monomer. The solid line connects results for an initial $\vert W_D\rangle$ electronic state while a dashed line is used for an initial $\vert \mathcal{E}^1_D\rangle$ state. Lines are only a guide for the eye.
    {\bf(b)} Population dynamics for the case of five qubits per monomer. Dashed lines represent results for $\vert \mathcal{E}^1_D\rangle$ as the initial electronic state, while solid lines are used when $\vert W_D\rangle$ is taken as the initial state. Orange, pink, and purple are used to signal the population of the donor sites, acceptor sites, and the W state in the acceptor $\vert W_A\rangle$, respectively. 
    {\bf(c)} Transfer rate $k_T$ as a function of the number of qubits ($N$), where the number of monomers, in this case, is given by $N/2$. The inset shows a schematic representation of how more monomers are incorporated into the system. Orange circles denote the simulation of the full Hamiltonian, while pink diamonds represent the simulation considering only the coupling between nearest-neighboring sites (see main text). Dashed and solid lines connecting markers are a guide for the eye. 
    {\bf(d)} Population dynamics for the case of five monomers. The population of the donor and the acceptor are identified by orange and purple solid lines, respectively, while the population of the three intermediate states $P_{I_1}$, $P_{I_2}$, and $P_{I_3}$ (second to fourth monomers) are indicated by different broken lines. $\vert T_D\rangle$ is the initial state for all results in panels (c) and (d). For all panels, we consider the parameter values of the parameter set $p_1$ (see Table ~\ref{table:parameter sets}).}
    \label{Fig5}
\end{figure*}
Similar to the $L=2$ case studied above, we can diagonalize the unperturbed Hamiltonian in the donor site while ignoring the inter-monomer couplings and obtain the following eigenstates for the $p=1$ case:
\begin{eqnarray} \label{epsiloeigenstates}
    &&\vert \mathcal{E}^1_D \rangle = \frac{1}{\mathcal{N}_1}\left(\vert 1 \rangle + \frac{2 \left(\sqrt{33}+3\right)}{\sqrt{33}+9}\vert 2 \rangle + \vert 3 \rangle\right), \nonumber \\
    &&\vert \mathcal{E}^2_D \rangle = \frac{1}{\mathcal{N}_2}\left(\vert 1 \rangle + \frac{2 \left(\sqrt{33}-3\right)}{\sqrt{33}-9}\vert 2 \rangle + \vert 3 \rangle\right), \nonumber \\
    &&\vert \mathcal{E}^3_D \rangle = \frac{1}{\sqrt{2}}(\vert 1 \rangle - \vert 3 \rangle),\;\;\;\;
\end{eqnarray}
where $\mathcal{N}_1$ and $\mathcal{N}_2$ are normalization constants (see Appendix \ref{app_monomer}). Similarly, acceptor states $\vert \mathcal{E}^1_A \rangle$, $\vert \mathcal{E}^2_A \rangle$, and $\vert \mathcal{E}^3_A \rangle$ can be equivalently expressed in terms of the states $\vert j_A \rangle$. We can write the Hamiltonian in Eq.~\eqref{eq_H_model} in this basis (see Appendix \ref{app_monomer} for details). The resulting couplings can be written in the notation $J_{AB}$ where $A,B = \mathcal{E}^1, \mathcal{E}^2, \mathcal{E}^3$.

As shown explicitly in Appendix \ref{app_monomer}, the larger of the inter-monomer couplings for $p=1$ is $J_{\mathcal{E}^1 \mathcal{E}^1}$. Importantly, the state $\vert \mathcal{E}^1_D \rangle$ has a large overlap with the $W$ state $\vert W_D\rangle = \frac{1}{\sqrt{3}}(\vert 1 \rangle + \vert 2 \rangle + \vert 3\rangle)$. 
Similar to the $L=2$ case studied previously,
the transfer is also favored by an initial symmetric superposition of product states if $J_{ij}>0$. 
Since the $W$ state is a paradigmatic state in quantum information science \cite{Dur2000} that can be efficiently prepared with digital protocols \cite{Cruz2019}, we study how transfer from an initial $\vert W_D\rangle$ state behaves in comparison to the initial state $\vert \mathcal{E}^1_D\rangle$. Specifically, in Fig.~\ref{Fig5}(a-b), we study how the transfer rate changes as the number of qubits per monomer $L$ is increased. The number of eigenstates $\vert \mathcal{E}^{j_D}_D\rangle$ increases linearly with the number of qubits in each monomer $L$. Here, for consistency, $\vert \mathcal{E}^1_D \rangle$ is defined as the state with the largest overlap to the corresponding $W$ state for each $L$. As shown in Appendix \ref{app_monomer}, this state has the highest eigenenergy and always yields the largest inter-monomer coupling strength for any monomer size $L$ considered here. 

In Fig.~\ref{Fig5}(a), we show the dependence of the transfer rate defined in Eq.~\eqref{transferrate} on the number of qubits per monomer $L$. As expected, for larger system sizes, the transfer rate decreases. For each $L$, the transfer from the initial state $\vert \mathcal{E}^1_D \rangle$ is always comparable to the $W$ state because of the large overlap of these two states. However, since the fastest processes favor the transfer to the $\vert \mathcal{E}_A^1\rangle$ state (see Appendix \ref{app_monomer}), the final state of the system will not be the $\vert W_A \rangle$ state, but it will be closer to $\vert \mathcal{E}_A^1\rangle$, even when the system is initialized in $\vert W_D \rangle$, as is illustrated in the dynamics depicted in Fig.~\ref{Fig5}(b). \crr{Given that the system is initialized in the entangled state $\vert W_A \rangle$ and evolves almost completely into the state $\vert \mathcal{E}_A^1 \rangle$, a large portion of the entanglement is transferred from donor to acceptor, similar to the case of simpler monomers (see Appendix \ref{app_entanglement} for more details). }


%

\subsection{Beyond Two Monomers}
Finally, we consider the case where the number of monomers is increased while the number of qubits per monomer is kept fixed to $L=2$, as shown in the inset of Fig.~\ref{Fig5}(c). In the case of $N=6$ (three monomers), the product basis is the same as for the $L=3$ case, namely, $\vert j \rangle$ with $j=1,\dots,6$. As each monomer is composed of only two qubits, the triplet-singlet basis is still the eigenbasis of the unperturbed Hamiltonian with:
\begin{eqnarray}
    &&\vert T_D \rangle = \frac{1}{\sqrt{2}}(\vert 1 \rangle + \vert 2 \rangle), \vert S_D \rangle = \frac{1}{\sqrt{2}}(\vert 1 \rangle - \vert 2 \rangle ),\;\;\;\; \nonumber \\
    &&\vert T_I \rangle = \frac{1}{\sqrt{2}}(\vert 3 \rangle + \vert 4 \rangle), \vert S_I \rangle = \frac{1}{\sqrt{2}}(\vert 3 \rangle - \vert 4 \rangle ),\;\;\;\; \nonumber \\
    &&\vert T_A \rangle = \frac{1}{\sqrt{2}}(\vert 5 \rangle + \vert 6 \rangle), \vert S_A \rangle = \frac{1}{\sqrt{2}}(\vert 5 \rangle - \vert 6 \rangle ).\;\;\;\;
\end{eqnarray}
Here, the label $I$ is used for the intermediate site. As we are interested in irreversible dynamics to the acceptor, we require the energy shift on the acceptor sites to be the lowest. Consequently, we choose a tilted energy configuration $\epsilon_1=\epsilon_2 = -\epsilon_3 = -\epsilon_4 = -\frac{1}{3}\epsilon_5 = -\frac{1}{3}\epsilon_6=\epsilon$. For the spin-phonon coupling strengths, we choose the alternating sign configuration $g_1=g_2=-g_3=-g_4=g_5=g_6=g$, which ensures that neighboring sites have non-zero Franck-Condon factor (see Eq.~\eqref{eq_FGR} and the text below). We keep this same prescription for larger values of $N$, where qubits $1$ and $2$ always correspond to the donor site, while qubits $N-1$ and $N$ correspond to the acceptor site, and all other qubits correspond to different intermediate sites. An alternative configuration is considered in Ref. \cite{Myers2015vaenergypumps}, where the spin-phonon couplings are decreasing, and the energies are staggered.

The excitation dynamics in Fig.~\ref{Fig5}(d) shows that negligible population is trapped in the intermediate states $\ket{T_{I_i}}$. This happens because a chain of irreversible transfers occurs between $\ket{T_{I_i},n}$ and $\ket{T_{I_{i+1}},n'}$, with $n'<n$ due to the interplay between coherent transfer and dissipation. The process is iterated until the population is trapped in state $\ket{T_A,0}$ for zero or very low temperature.
We shall note that although any triplet state in one site is resonant with multiple triplet states with different numbers of phononic excitations in subsequent sites, these processes are suppressed by the power-law decay coupling strengths and the Franck-Condon factors. This setting makes the interaction between neighboring sites the dominant process. We show in Fig. \ref{Fig5}(c) that the change of the transfer rate is negligible when setting $J_{ij}=J/d_{ij}^p=0$ for $d_{ij}>5$ (see inset), which eliminates the next-nearest-neighbor couplings between monomers. However, even in this case, long-range interactions between different qubits in neighboring monomers still play a crucial role in favoring the transfer of delocalized states. It is important to note that for $N>4$, it is necessary to substitute the donor population $P_D$ by $1-P_A$ in the definition of transfer rate in Eq.~\eqref{transferrate}.

As described above, in the absence of disorder, the dynamics for multiple monomers involves predominantly the triplet states in each site, given that the system is initialized in $\vert \psi_{T_D,0} \rangle$. In this case, the dynamics reduces to the quantum state transfer from donor site $1$ to acceptor site $N/2$ in a chain of $N/2$ sites made of $N$ qubit ions. Recently, a bound for the optimal quantum state transfer was found in the case of a spin chain with power-law decaying interactions and conservation of the number of excitations \cite{Jameson_2024}. While the model presented here fulfills those two conditions, it also considers the spatially inhomogeneous spin-phonon interaction and the dissipation induced by the bath. 
The time to reach the steady state (see Fig.\ref{Fig5}(d), for example) is always considerably larger than the optimal time $t_{\rm opt} = {\pi}/{J\sqrt{N}}$ reported in Ref. \cite{Jameson_2024}. An interesting open question is whether a state transfer tight bound can be found for power-law systems featuring spin-phonon coupling and dissipation. 



\section{Experimental Design}\label{sec_ions}
In this section, we show how the master equation in Eq.~\eqref{eq_master} can be realized natively in a linear trapped-ion chain consisting of $N$ site ions and $M$ coolant ions with state-of-the-art experimental techniques. We also provide realistic experimental parameters that would allow the analog quantum simulation of the transfer dynamics studied in the previous sections with precise control over the relevant system parameters.

\subsection{Trapped-Ion Model of Excitation Transfer} \label{sec_ions1}

Two long-lived internal states with a frequency difference $\omega_0$ can be used to encode the electronic state of the site ions, and thus, the $N$-qubit state represents the electronic configuration of the system that can be manipulated with coherent Raman drives. At the same time, the $M$ coolant ions can be employed to engineer the reservoir for the collective bosonic mode via sympathetic cooling by using ground-state cooling with either Raman \cite{Monroe1995} or resolved sideband one-photon transitions \cite{Cetina2022,so2024trappedion}. For example, the setup can be operated with either a chain of $N+M$ $^{171}$Yb$^+$ ions, where the laser beam arrangement assigns the $N$ qubit and $M$ coolant ions, or a $^{171}$Yb$^+$-$^{172}$Yb$^+$ dual-species crystal, where the $^{171}$Yb$^+$ ions encode the electronic state of the system, and the $^{172}$Yb$^+$ ions undergo cooling \cite{so2024trappedion}. We note that the cooling efficiency relies on the degrees of participation between the coolant ions and the motional mode that encodes the vibrational degree of freedom in Eq.~\eqref{eq_H_model}. 
For this reason, an even-numbered $N+M$ ion chain is preferred because all the ions would have non-vanishing participation in all the collective modes of the crystal due to its mirror symmetry, which maximizes the number of modes available for vibrational mode encoding and reservoir engineering \cite{Hou2024cooling}.


Individual Raman beams with multiple frequency beatnotes of different amplitudes and phases on the qubit ions can be used to generate the desired Hamiltonian in Eq.~\eqref{eq_H_model}. A frequency beatnote on resonance with the qubit frequency $\omega_0$ (carrier drive) allows for individual energy shifts to modulate the energy landscape of the system, $\sum_{i}\sigma_i^z\epsilon_i/2$, in the appropriate spin basis. The site-to-site electronic coupling term, $\sum_{i< j} J_{ij}( \sigma_i^{+} \sigma_j^{-}+ \sigma_i^{-} \sigma_j^{+})$, can be realized natively by simultaneously driving two symmetric beatnotes $\omega_0\pm(\omega_N+\delta_J)$, where $\omega_N$ is the secular frequency of the center-of-mass collective mode. This results in a long-range Ising XX interaction with the coupling strength $J_{ij}=J/d_{ij}^p$, where $p$ can be adjusted by changing the axial frequency of the ion chain and the detuning $\delta_J$. \crr{The tunable range of $p$ is between 0 and 3 \cite{monroe2021programmable}, but the experimentally accessible range is approximately between 0.5 and 2.} A strong transverse field $B\sum_i\sigma^z_i$ ($J\ll B \ll \delta_J$) can be applied to suppress the double excitation processes present in the native Ising interaction, therefore enforcing the conservation of the total number of spin excitations along the transverse field direction. 
This can be realized either by shifting the two symmetric sideband frequency tones used for generating the Ising interaction with a common detuning or by implementing an additional qubit frequency resonant tone, depending on the spin basis chosen for the Hamiltonian engineering \cite{monroe2021programmable}. In the following section, which describes an experimental setup for simulating the two-monomer case, we demonstrate the use of the latter approach.
The residual $B$ field only adds a global offset to the energy landscape of the system and does not affect its dynamics. 

Similarly, both the vibrational and vibronic coupling terms of the model, $\omega a^\dagger a + \sum_i ({g_i}/{2})\sigma_i^z(a^\dagger +a)$, can be engineered with another pair of symmetric beatnotes $\omega_0\pm(\omega_k-\omega)$, where $\omega_k$ is the secular frequency of a chosen reservoir collective motional mode described by $a^\dag (a)$ operators, and $\omega$ is the detuning from the reservoir motional mode with $\omega\sim g_i$ \cite{Schneider_2012, so2024trappedion}. 

\begin{figure*}[t!]
\includegraphics[width=0.8\textwidth]{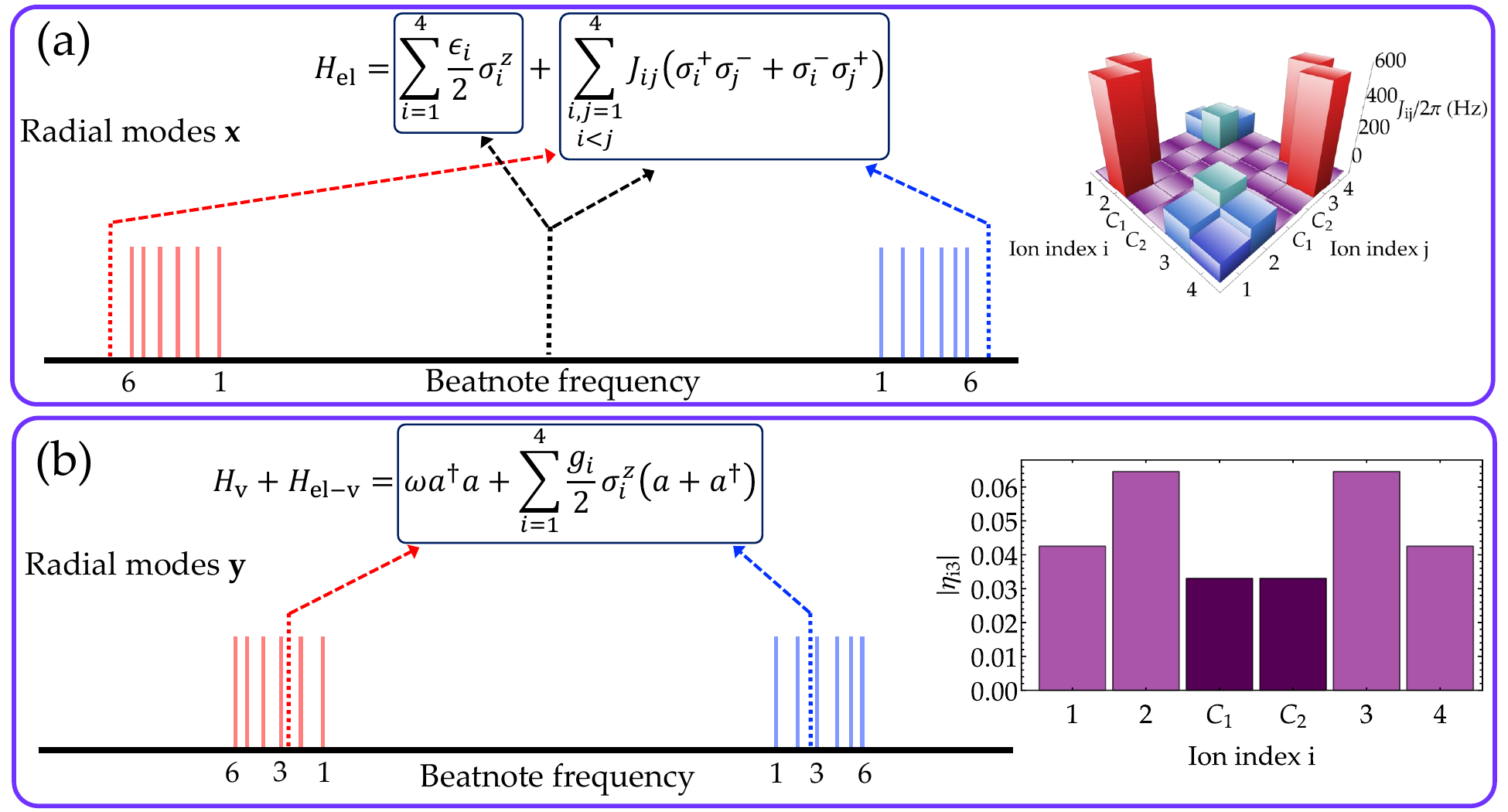}
\caption{\textbf{Experimental configuration for simulating the two-monomer model with a chain of six $^{171}$Yb$^+$ ions:} {\bf(a)} Multi-frequency drives on the qubit ions (represented by labels 1, 2, 3, and 4) to generate the electronic terms $H_\text{el}$ of the system Hamiltonian and the corresponding $J_{ij}$ matrix with $J/2\pi=0.62$ kHz and $p\approx1$. {\bf(b)} Bichromatic drive on the qubit ions to generate the vibrational $H_\text{v}$ and vibronic coupling $H_\text{el-v}$ terms using the 3rd motional mode. The solid (dashed) vertical lines represent the motional sidebands (355 nm laser beatnote frequency tones). The $J_{ij}$ matrix and Lamb-Dicke parameters $\eta_{i3}$ are calculated with the trapping condition, where the center-of-mass mode frequencies along the three principal axes are $\{\omega_6^x,\omega_6^y,\omega_1^z\}=2\pi\times\{3.2, 2.8, 0.5\}$ MHz. Simultaneously, the two central coolant ions (denoted by $\text{C}_1$ and $\text{C}_2$) undergo continuous resolved sideband cooling on the 3rd motional mode of the radial modes $y$ to emulate the dissipation process of the vibrational mode of the two-monomer system.}
\label{fig:exp_setup}
\end{figure*}

It shall be noted that if a single set of motional modes were used for both bichromatic drives, the approximate effective Hamiltonian stemming from the Magnus expansion of the time-ordering evolution operator would give rise to higher-order, non-commuting electronic coupling terms \cite{Davoudi2020}. To mitigate this problem, we propose to use orthogonal sets of motional modes to generate the two non-commuting parts of the Hamiltonian separately \cite{deng2005effective, Zhu2023Pairwise}: the electronic coupling interactions can be generated by driving the higher frequency radial mode set along the $x$ principal axis of the trap, and the vibronic coupling interactions can be realized by using one mode from the lower frequency radial mode set along the $y$ principal axis of the trap (see Fig.~\ref{fig:exp_setup}) since it is standard practice in trapped-ion systems to break the degeneracy of the radial modes into the higher and lower frequency sets by applying a radial static quadrupole potential upon the radial confinement potential induced by the radiofrequency drive. This choice has the advantage of ensuring spectral separation between the two drives, reducing crosstalks, and resulting in two drives having very different detunings with respect to their respective normal modes.

We also note that the proposed scheme to generate the vibronic coupling and the harmonic terms can off-resonantly couple to other collective motional modes $(m\neq k)$ and induce additional two-body interactions. However, these terms result only in state-dependent shifts to the energy landscape of the system and do not affect its single electronic excitation manifold description. These energy shifts can be corrected experimentally by adjusting the single-site energy offsets (see Appendix \ref{app_crosstalk}).

\crr{Finally, the simultaneous application of the four sideband tones in both bichromatic drives can induce A.C. Stark shifts, which introduce spurious terms into the engineered trapped-ion Hamiltonian, deviating away from the desired model. Therefore, in practice, one also must compensate for these shifts with a proper choice of the amplitudes and frequencies of the laser tones, which have been successfully achieved by several analog trapped-ion experiments with a similar setup using two simultaneous bichromatic tones \cite{Valahu2023,so2025quantumsimulationschargeexciton,navickas2024experimentalquantumsimulationchemical}.} 
\begin{table}[h!]
\centering
\begin{tabular}{lll}
\toprule
\textbf{System parameter} & \medspace \textbf{Experimental range} \\
\text{($\hbar = 1$)} & \medspace \text{$[2\pi \times ~\text{kHz}]$}\\
\midrule
Local energy shift $\epsilon_i$ & \medspace $|\epsilon_i|\le 100$ \\ 
Site-to-site electronic & \medspace $\lesssim 1$ with $0.5\lesssim p \lesssim 2$ \\
coupling $J_{ij}=\frac{J}{d_{ij}^p}$\\
Vibrational energy $\omega$ & \medspace $2\text{--}40$\\ 
Vibronic coupling $g_i$ & \medspace $\lesssim 40$\\ 
Dissipation rate $\gamma$ & \medspace $\lesssim 5$\\ 
Average phonon number $\bar{n}$  \medspace  \medspace & \medspace $\gtrsim 0.01$\\ 
\bottomrule
\end{tabular}
\caption{Typical ranges for the relevant system parameters accessible by an analog trapped-ion quatum simulator.}
\label{table:exprange}
\end{table}

Combining these drives leads to a trapped-ion Hamiltonian that directly maps to the model in Eqs.~\eqref{eq_H_model} with individual control knobs on the system parameters, $\epsilon_i,\; g_i,\; J_{ij},$ and $\omega$, through the optical power, frequency, and phase of the individual-addressing Raman beams. Likewise, the local implementation of continuous resolved sideband cooling of the collective motional mode associated with $a^\dagger (a)$ operators on the $M$ coolant ions generates the dissipator in Eq.~\eqref{eq_master} with tunable dissipation rate {$\gamma$} and bath temperature $\bar{n}$ via the optical powers and frequency detunings of the cooling laser beams \cite{MacDonell2021simulation, Lemmer2018, so2024trappedion}. Table~\ref{table:exprange} summarizes the experimental ranges for the relevant system parameters accessible by a standard analog trapped-ion quantum simulator, which allows vast combinations of parameters for extensive studies of the model discussed above.

\subsection{Two-monomer model} \label{sec_ions2}
The experimental demonstration of our results on the two-monomer case in Sec. \ref{two_monomer_sec} can be realized with the setup in Fig.~\ref{fig:exp_setup}. The simulator consists of six $^{171}$Yb$^+$ ions trapped at the center-of-mass secular frequencies of $\{\omega_6^x,\omega_6^y,\omega_1^z\}=2\pi\times\{3.2, 2.8, 0.5\}$ MHz along the three principal axes, where $z$ is the axial direction of the trap. We shall note that $\alpha=x,\;y,\;z$ in $\omega_m^\alpha$ is different from $\beta$ in the operator $\sigma_i^\beta$, which describes the spin basis in the Bloch sphere. This trapping configuration would provide an average spacing between the neighboring ions of approximately $3.50\;\mu$m.
In the driven rotating frame, where two $\pi/2$ pulses are employed to map the $z$ spin basis of Eq.~\eqref{eq_H_model} onto the $y$ basis of the trapped-ion Hamiltonian, the combination of a bichromatic 355 nm Raman drive with the frequency beatnotes of $\omega_0\pm(\omega_6^x+\delta_J)$ and a carrier drive at the beatnote frequency of $\omega_0$ on the qubit ions induces a power-law decay Ising Hamiltonian with a transverse field, $\sum_{i<j}J_{ij}\sigma_i^x\sigma_j^x+\sum_i B\sigma_i^z$, where $\omega_0$ is the frequency difference between the clock hyperfine states of $^{171}$Yb$^+$ that represent the qubit, and $J_{ij}=J/d_{ij}^p$. For the experimentally feasible values of $\{J,\;B,\;\delta_J\}=2\pi\times\{0.62,\;6.4,\;66\}$ 
kHz, the condition $J\ll B \ll \delta_J$ is sufficiently satisfied resulting in the spin-hopping interactions $\sum_{i< j} J_{ij}( \sigma_i^{+} \sigma_j^{-}+ \sigma_i^{-} \sigma_j^{+})$ with $p\approx1$. Moreover, another carrier drive is used to generate the energy landscape of the system, and the combined interactions become the electronic terms of the two-monomer Hamiltonian.

As mentioned above, using a suitable mode from the other radial mode set to generate the vibrational and vibronic coupling terms eliminates higher-order, non-commutating electronic coupling terms from the Magnus expansion of the time-evolution operator. To realize our results, it is important to balance the vibronic coupling strengths ($|g_i| = g$) among the qubit ions. Since $|g_i|=|\eta_{ik}|\Omega_i$, where $\Omega_i$ is the Rabi coupling strength to ion $i$ and depends on the laser beam power, the choice of mode $k$ must consider the overall absolute differences among the Lamb-Dicke parameters $|\eta_{ik}|$ of the qubit ions, which determine the degree of participation of the ion $i$ to the motional mode $k$. Hence, the required laser powers to compensate for the different couplings are reduced by choosing a mode $k$ that has the least differences among the participation coefficients $|\eta_{ik}|$. A more balanced participation of the ions to the mode of choice also leads to a more efficient sideband cooling on the coolant ions. Although the center-of-mass mode seems like an obvious choice for this criteria due to its intrinsic balanced participation coefficients $|\eta_{i6}|=\eta/\sqrt{6}$, it is the most sensitive mode to electric field noise that typically limits the motional coherence time of trapped-ion experiments. Therefore, in the case of six ions, the best option for this scheme is mode $3$ (see Fig.~\ref{fig:exp_setup}(b)) that can be driven with a bichromatic 355 nm Raman drive with the frequency beatnotes of $\omega_0\pm(\omega_3^y-\omega)$, appropriate phases, and offset-corrected laser powers to generate the vibrational and vibronic coupling terms of the two-monomer Hamiltonian with realistically achievable $\omega = g = |\eta_{i3}|\Omega_i = 2\pi\times$ 2 kHz.

Throughout the simulation, it is possible to use individual laser cooling beams onto the two central ions to continuously remove the system's vibrational excitations. This can be achieved, for example, via continuous resolved sideband cooling on the 3rd motional mode of the $y$-direction of the trap axes using the narrow-linewidth optical transition of $^2S_{1/2} \leftrightarrow$ $ ^2D_{5/2}$ at 411 nm along with a repumper at 1650 nm \cite{vybornyi2023sideband}.



In conclusion, owing to the tunability of the trapped-ion system, it will also be possible to investigate the Frenkel exciton model studied in this paper experimentally by singling out the individual roles of the system parameters at the site level ($\epsilon_i,\; g_i,\; J_{ij},$ and $\omega$), bath properties ($\gamma$ and $\bar{n}$), and different electronic configurations in governing the transfer dynamics over wide ranges of parameters. The extension to longer chains with multiple intermediate sites or more complex monomer structures in our proposed setup can be achieved by adding more ions and more coolants. For long chains, a combination of quadratic and quartic axial potentials \cite{Lin2009anharmonic} can be used to achieve quasi-uniform ion spacing decreasing crosstalk in addressing and detection. For example, a quantum simulator of 28 $^{171}$Yb$^+$ ions can be used to simulate six monomers composed of three qubit ions each and interleaved with two coolant ions.
The center-of-mass secular frequencies can be set to $\{\omega_{28}^x,\omega_{28}^y,\omega_1^z\}=2\pi\times\{3.2, 2.8, 0.238\}$ MHz leading to an average spacing between the neighboring ions of approximately $3.50\;\mu$m, similar to the proposed setup for simulating the two-monomer model. With appropriate laser powers, the system parameter values of $\{J,\;B,\;\delta_J,\; p,\;\omega,\;g\}=2\pi\times\{0.58,\;5.38,\;50,\;1.28,\;2,\;2\}$ kHz are experimentally accessible for realizing the dynamics of vibrationally assisted exciton transfer in a six-monomer system.

As an alternative strategy to scale up to larger systems, it is possible to use qudits instead of qubits, where we encode the monomer's electronic configuration with more than two internal states of the ions \cite{MacDonell2021simulation}. However, this comes at the cost of engineering complex electronic interactions with additional lasers to create the desired Hamiltonian. 

Lastly, a key component
of the simulation of more complex monomers
proposed here is the preparation of $W$ states (or states with a large overlap with $W$ states), which is achievable through digital \cite{Cruz2019} or analog methods \cite{Zhu2025}. This hybrid digital-analog approach is also ideal for implementing the recently proposed simulation of 2D spectroscopy using a probe qubit \cite{guimarães2024acceleratingtwodimensionalelectronicspectroscopy}, which significantly reduces measurement overhead compared to standard digital protocols \cite{Lee2021}.



\section{Discussion and Outlook}\label{sec_outlook}

In this work, we characterized the role of entanglement and delocalization in long-range interacting spin chains interacting with an engineered reservoir. We found that the spin-spin interaction defines a specific class of entangled states that optimize the transport of excitations in the presence of an engineered environment. This conclusion still holds when accounting for static disorder, \crr{noise}, and thermal excitations. 

Our work is motivated by very recent experimental advances \cite{Gorman2018, Sun2023,sun2024quantumsimulation, so2024trappedion, navickas2024experimentalquantumsimulationchemical} and aims at providing a blueprint for the analog quantum simulation of Frenkel exciton models based on trapped-ion hardware. Our work is a step towards the direct trapped-ion simulation of vibrationally assisted exciton transfer in chemical and biological complexes, where the role of quantum coherence and delocalization in electron transfer is still debated \cite{Engel2007, Plenio2013, Monahan2015, Romero2014, Fassioli2014, Arsenault2020, Sneyd2021, Sneyd2022}. A trapped-ion quantum simulator offers, on the one hand, full tunability of both the coherent and dissipative parts of the evolution while, on the other hand, granting access to site-resolved observables that are crucial to understanding the energy flow dynamics \cite{Monahan2015}. 

Trapped-ion quantum simulators can naturally access the intermediate coupling regime \cite{Fassioli2014}, where the reorganization energy $\lambda$ is of the same order or larger than the electronic coupling between different molecular sites. This regime poses significant computational challenges for classical methods, such as tensor networks, especially when considering long-range interactions \cite{Somoza2019, Kang2024}.


In this work, we focused on a spin chain coupled to a single, damped phonon mode. However, electron transfer in realistic systems often involves multiple vibrational modes.
Therefore, the inclusion of multiple damped modes of motion will be paramount for the simulation of quantum dynamics of complex excitonic models. This can be achieved natively on trapped-ion quantum hardware, where multiple coolant ions can be used to control the individual cooling rates and the temperature of multiple bosonic modes. This capability supports the engineering of structured spectral densities \cite{Lemmer2018}, paving the way for studying non-Markovian effects in quantum dynamics \cite{Wang2024simulating,debecker2024spectral, debecker2024controlling} \crr{where strong, ultra-strong, and time-dependent system-environment couplings can be studied}. Furthermore, the realization of Frenkel exciton models with multiple modes with engineered spectral density will allow simulation across diverse vibrational frequency ranges, capturing both inter- and intra-molecular vibrations and low-frequency solvent environments \cite{Plenio2013, Tiwari2013}.

Including more than one bosonic mode in this model will also allow the simulation of multiply bridged donor-acceptor molecules \cite{Goldsmith2006}, where intermediate sites experience distinct environments or couplings to different vibrational modes \cite{Li_2021Whaley,Wu2016}. Additionally, this setup provides an opportunity to introduce tunable anharmonic couplings among bosonic modes in the trapped-ion chain \cite{Ding2017}, facilitating the study of anharmonic effects, which are believed to influence electron transfer \cite{Zhang2023pnas}.  

\crr{In the work presented here, we focused on the role that coherent superpositions of states within each monomer play in determining the transfer efficiency. Considering more than two monomers, as we did in Section~\ref{sec_trimers}, opens the possibility of studying the transfer of an excitation initially delocalized over multiple monomers. The presence of different types of quantum coherences and their survival over time has been found to be related to the high efficiency of excitation transfer in photosynthetic complexes \cite{Engel2007,Lee2009coherence}. Previous studies  \cite{Jang2008,Ishizaki2009Adequacy,Ishizaki2009Unified,Ishizaki2009} have pointed out that the consideration of non-Markovian baths is essential to model correctly the survival of these coherences, which, as pointed out earlier, represents an interesting future research direction both experimentally and theoretically.}

Here, we explored the effect of static disorder \crr{and white-noise fluctuations of the energy shifts through electronic dephasing}, \crr{the latter as well as} motional dephasing are \crr{common noise sources in realistic setups and play a key role in transfer dynamics} \cite{Caruso2009, Li_2022Whaley,Li_2024Whaley}. In trapped-ion systems, electronic dephasing can be mimicked by introducing noisy light shifts \cite{Maier2019, MacDonell2021simulation}, and motional dephasing can be realized by controlling the trap frequencies \cite{olayaagudelo2024simulatingopensystemmoleculardynamics}.
\crr{As shown in Ref.~\cite{Maier2019}, more complicated noise profiles can be realized in these platforms, opening the possibility to study richer coherence dynamics.}
While our study focuses on excitation-conserving~$\sim\sigma_z(a^\dag+ a)$ spin-phonon coupling, real systems feature relaxation of excitations, which limits the transfer efficiency. In trapped-ion systems, it is possible to include spin relaxation experimentally by using a red-sideband (Jaynes-Cummings type) drive with dissipation.

A further extension of our model involves generalizing qubits to qudits, allowing the simulation of other transfer phenomena such as singlet fission processes \cite{Collins2023, campaioli2024optimisationultrafastsingletfission}. Trapped-ion systems allow coherent coupling of multiple atomic states to bosonic modes. Spin-1 Hamiltonians, for example, have been implemented with trapped ions \cite{Senko2015}, and qudit-based gates have been realized \cite{edmunds2024constructingspin1haldanephase}. 

From a quantum information perspective, our model raises intriguing questions about generalizing recently established bounds on quantum state transfer in long-range interacting systems \cite{Sweke_2019,Guo2020, Tran2021, Jameson_2024} to open quantum systems with vibronic couplings. Such bounds could help optimize state transfer across the large parameter space $(J_{ij}, \epsilon_i, g_i, \gamma)$ inherent in these models. 
Exploring larger collections of monomers with intermediate coupling, long-range interactions, and dissipation could also illuminate parameter regimes where classical methods are computationally prohibitive, offering insights into the design of light-harvesting materials.
\newpage
\begin{acknowledgments}
We acknowledge Francesco Campaioli and Peter G. Wolynes for insightful comments. G.P. acknowledges the support of the Welch Foundation Award C-2154, the Office of Naval Research Young Investigator Program (Grant No. N00014-22-1-2282), the NSF CAREER Award (Grant No. PHY-2144910), the Army Research Office (W911NF22C0012), and the Office of Naval Research (Grants No. N00014-23-1-2665 and N00014-24-12593). We acknowledge that this material is based on work supported by the U.S Department of Energy, Office of Science, Office of Nuclear Physics under the Early Career Award No. DE-SC0023806. H.P. acknowledges support from the NSF (Grant No. PHY-2207283) and from the Welch Foundation (Grant No. C-1669).


\end{acknowledgments}

\bibliography{Dimer}

\appendix

\section{Model}\label{app_A}

To find the eigenstates of the Hamiltonian in Eq.~\eqref{eq_bosonic_Ham} for a given $j$, we displace the bosonic mode as $a = \alpha_{j} + \tilde{a}_j$, where $\alpha_{j}$ is a complex number and $\tilde{a}_j$ is another bosonic annihilation operator. In terms of the new variables:

\begin{eqnarray}
    H_{B,j} &=&\omega \tilde{a}^{\dagger}_j \tilde{a}_j + \omega (\alpha_{j})^2 + \omega \alpha_j (\tilde{a}_j+\tilde{a}^{\dagger}_j) \nonumber \\
    &&+ g_j \alpha_j + \frac{g_j}{2}(\tilde{a}_j+\tilde{a}^{\dagger}_j)\,.
\end{eqnarray}

To eliminate the linear terms on the annihilation and creation operators, we set $\alpha_j = \frac{-g_j}{4 \omega}$. The Hamiltonian becomes:

\begin{equation}
    H_{B,j} = \omega \tilde{a}^\dagger_j \tilde{a}_j - \frac{g_j^2}{4 \omega}\, \label{eq_diagonal_bos_pert}
\end{equation}

corresponding to a harmonic oscillator with a constant energy shift caused by the displacement of the oscillator by $\alpha_j = - \frac{g_j}{2\omega}$. The eigenstates of Eq.~\eqref{eq_bosonic_Ham} are known as displaced Fock states \cite{Wunsche_1991} and can be represented by $D(\alpha_j)\vert n\rangle = \vert n, \alpha_j \rangle$, where $n$ is the number of excitations in the displaced oscillator and $D(\alpha)$ is the bosonic displace operator. The corresponding eigenvalues are given by $n \omega - g_j^2/(4\omega)$. Consequently, the eigenstates of the full unperturbed Hamiltonian in Eq.\eqref{H01} are given by $\ket{\psi_{j,n}} = \ket{j}\otimes \ket{n,\alpha_j}$ with eigenergies given by Eq.\eqref{eq_E_0}.

To derive the expression for the Fermi golden rule (FGR) in Eq.~\eqref{eq_FGR}, we start by considering an initial state $\vert \psi_{i,0}\rangle = \vert i \rangle \otimes \vert 0, \alpha_i \rangle$ characterized by zero bosonic excitations $n=0$ and an initial electronic state $\ket{i}$ (for most of our two-monomer results $\ket{i} = \ket{T_D}$), and final states of the form $\ket{\psi_{f,n}} = \ket{f} \otimes \ket{n,\alpha_f}$. The FGR is defined as:

\begin{equation} \label{FGR_definition}
    k_T^{if} = \sum_{n=0}^\infty \vert \langle \psi_{i,0} \vert \psi_{f,n}\rangle\vert^2 \frac{\gamma}{(E_{i,0}-E_{f,n})^2+\gamma^2/4}\,,
\end{equation}

where $\gamma$ has been incorporated through a Lorentzian broadening of the delta functions that appear in the conventional definition of the FGR. Here, $\vert \langle \psi_{i,0} \vert \psi_{f,n}\rangle\vert^2 = \vert J_{if}\vert^2 \vert \langle 0, \alpha_i \vert n \alpha_j \rangle\vert^2$, where $J_{if}$ is the corresponding inter-monomer coupling. In this case, the Franck-Condon factors can be simplified further:

\begin{eqnarray}
 \langle 0, \alpha_i \vert n \alpha_j \rangle &=& \langle 0\vert D(-\alpha_i)D(\alpha_f) \vert n \rangle \nonumber \\
 && = \langle 0, \alpha_i - \alpha_f \vert n \rangle \nonumber \\
 && = e^{-(\alpha_i-\alpha_f)^2/2} \frac{(\alpha_i-\alpha_f)^n}{\sqrt{n!}}\,,
\end{eqnarray}

where in the last line we have used the expansion of a displaced state in terms of the Fock basis and the orthonormality condition $\langle m \vert n\rangle = \delta_{nm}$. Using this simplified form of the Franck-Condon factor we obtain the expression in Eq.~\eqref{eq_FGR}.

To compare the analytical expression $k_T = \sum_{f \in A} k_T^{if}$ with numerical results, we define the transfer rate in Eq.~\eqref{transferrate}. 
To extract the transfer rate, we simulate the dynamics for a total time $\omega t_{\rm sim}/2\pi =100$, which means that, in practice, we use the expression:

\begin{equation}
k_T = \frac{\int_{t=0}^{t_{\rm sim}} P_D(t)dt}{\int_{t=0}^{t_{\rm sim}} t P_D(t)dt} - \frac{2}{t_{\rm sim}}\,,
\end{equation}
where the second term is a correction for considering a finite time $t_{\rm sim}$. 

\section{Couplings}\label{app_couplings}

In section~\ref{two_monomer_sec}, we introduced the expression for the inter-monomer couplings in the triplet-singlet basis in Eq.~\eqref{couplings_ts}. In that expression, we used the convention of considering two coolants (see discussion below Eq.~\eqref{eq_Jij}) which sets $d_{23}=3$. More generally, the couplings are given by:

\begin{eqnarray} \label{couplings_ts_full}
    &&J_{TT} = \left(\frac{1}{2}\frac{J}{(1+d)^{p}} + \frac{J}{(2+d)^p} + \frac{1}{2}\frac{J}{(3+d)^{p}} \right), \nonumber \\
    &&J_{SS} = \left(-\frac{1}{2}\frac{J}{(1+d)^{p}}+\frac{J}{(2+d)^{p}} -\frac{1}{2}\frac{J}{(3+d)^{p}} \right), \nonumber \\
    &&J_{TS}= -J_{ST} = \left(\frac{1}{2}\frac{J}{(1+d)^{p}} -\frac{1}{2}\frac{J}{(3+d)^{p}} \right)\,.
\end{eqnarray}

\begin{figure}[h!]
\includegraphics[width=0.45\textwidth]{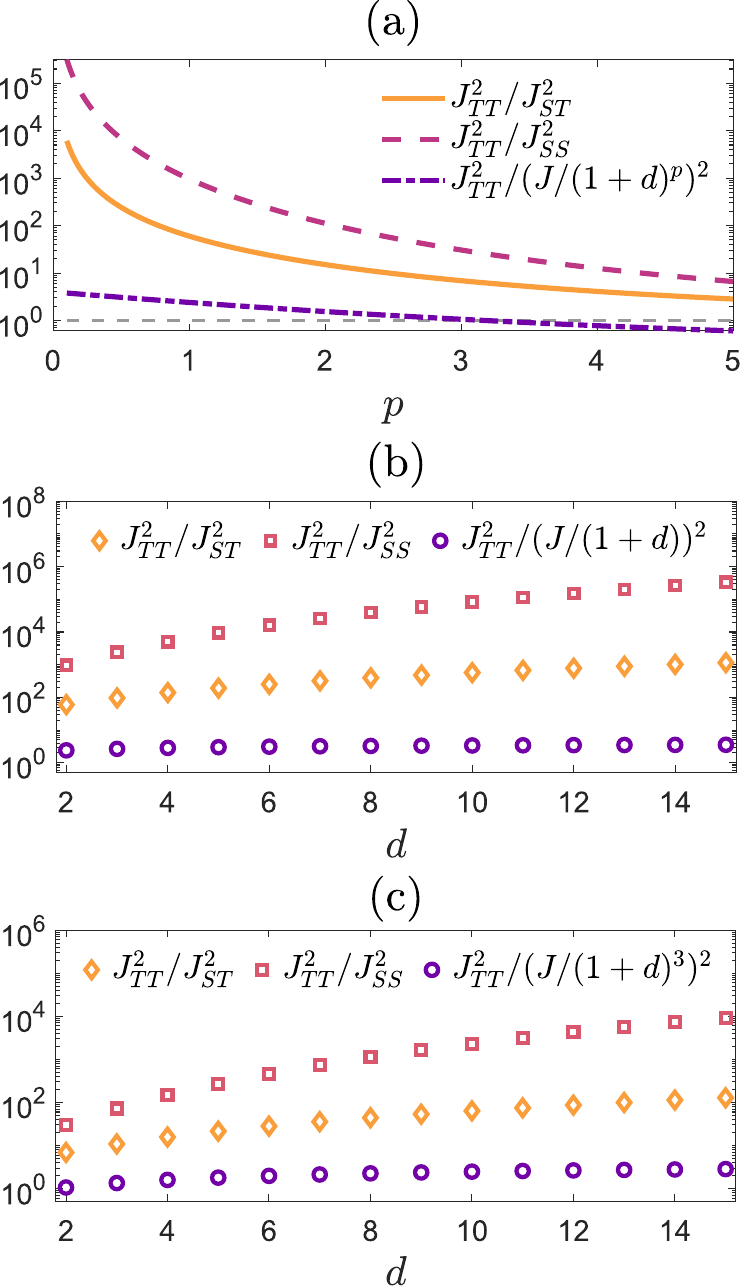}
\caption{{\bf Inter-monomer couplings in the triplet-singlet basis:} {\bf(a)} Ratio of the different inter-dimer couplings as a function of $p$ for $d=2$. The horizontal gray dashed line is used as a guide to highlight the point where $J_{TT}^2/(J/3^p)^2=1$ at $p \approx 3$.
{\bf(b)} Ratio of the different inter-dimer couplings as a function of $d$ for $p=1$.
\crr{{\bf(c)} Ratio of the different inter-dimer couplings as a function of $d$ for $p=3$.}}
\label{Fig1supp}
\end{figure}

Where the distance between neighboring ions is set equal to 1, and $d$ is the number of coolants. The ratio of the different couplings in Eq.~\eqref{couplings_ts_full} for different values of $p$ and $d$ is shown in Fig.~\ref{Fig1supp}.

The results reported in panel (a) show that $J_{TT}$ is the dominant process for all $p\lesssim 3$ and $d=2$. Around $p=3$, $J_{TT}$ becomes comparable to $J/(3^p)$, which is the largest inter-monomer coupling in the product basis. For all $p<3$, we would expect that transfer from the triplet state is faster than from any other product or entangled state.

For $p=1$, the results in Fig.~\ref{Fig1supp}(b) show that the ratio between $J_{TT}$ and any other coupling becomes larger as $d$ is increased. This means that increasing $d$ results in a more dominant triplet-triplet transfer, of course, at the cost of the transfer becoming slower. This trend is independent of the value of $p$. \crr{For example, the case $p=3$ representing the dipole interaction, which is native in many systems, shows a similar trend as a function of $d$, as shown in Fig.~\ref{Fig1supp}(c).}
In the main text, the chosen values of $p=1$ and $d=2$ ensure that the triplet-triplet transfer is the most dominant while still having a fast transfer. 
\crr{
\section{Numerical simulation details} \label{app_simulations}

Since the dynamics described in the excitation transfer involve a non-zero number of excitations in the phonon mode, especially in the case of higher temperatures, the numerical integration of Eq.~\eqref{eq_master} requires choosing an appropriate cutoff number of phonon excitations $N_{\rm ph}$. In Fig.~\ref{FigCutoff}, we show how the evolution of different observables is affected by the choice of this cutoff.

\begin{figure}[h!]
\includegraphics[width=0.45\textwidth]{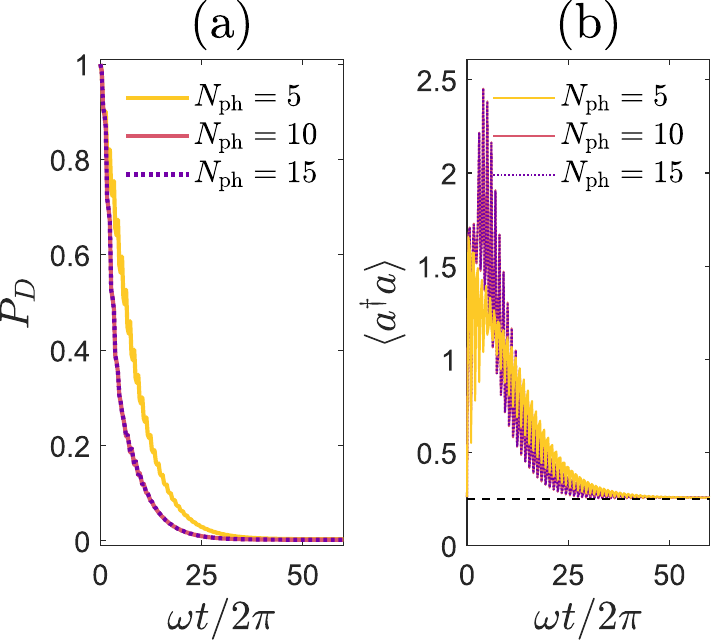}
\caption{{\bf Phonon cutoff dependence:} Time evolution of observables for different values of the phonon cutoff $N_{\rm ph}$, represented by distinct curves. The chosen parameters are the same as the parameter set $p_1$ in Table.~\ref{table:parameter sets}. 
{\bf(a)} Population in the donor as a function of time.
{\bf(b)} Population in the phonon mode as a function of time. The black dashed line represents $\alpha^2 = (g/2\omega)^2=0.25$, which is the expected value of the population in the steady state.} 
\label{FigCutoff}
\end{figure}

As depicted in the figure, for these parameters, a cutoff of $N_{\rm ph}>10$ excitations already provides an appropriate size of the phonon mode Hilbert space, as no significant changes can be noticed in the observables when the value is increased to $N_{\rm ph} = 15$ (the two curves lie on top of each other). We adjust the phonon cutoff accordingly for the different parameter sets used in this work.
}

\begin{figure}[t!]
\includegraphics[width=0.45\textwidth]{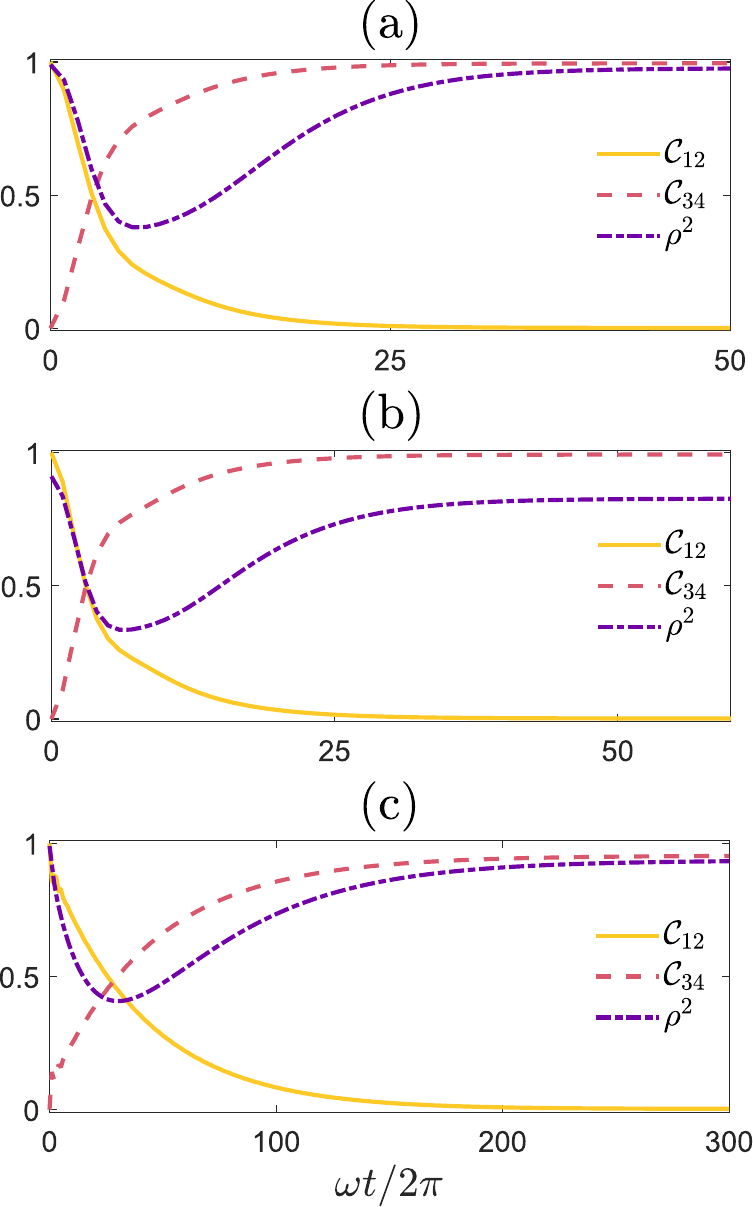}
\caption{{\bf Entanglement transfer:} Time evolution of the concurrence between different qubits and the purity of the state. 
{\bf(a)} The chosen parameters are the ones in parameter set $p_1$ in Table.~\ref{table:parameter sets}. 
{\bf(b)} Same as in (a) but with $\bar{n} = 0.1$.
{\bf(c)} Same as in (a) but with $\epsilon = 0.26 \omega$.} 
\label{FigEntanglement}
\end{figure}

\crr{
\section{Entanglement transfer} \label{app_entanglement}

One of the main results in this work is that the initialization of the system in an entangled (delocalized) state of the qubits in the donor site speeds up the transfer into the acceptor site. Under ideal conditions (i.e., at low temperature and on resonance), the steady state in the acceptor is also an entangled state, as illustrated in Fig.~\ref{Fig2}. For example, this is evidenced by the fact that $P_D \approx 1 - P_{T_A}$ when the system is initialized in the electronic entangled state $\vert T_D \rangle$. However, we make this argument more concrete here, where we explicitly show the dynamics of the qubit's entanglement in terms of the concurrence.

For a two-qubit state, pure or mixed, the concurrence can be used to determine the entanglement. Mathematically, the concurrence is defined as \cite{bennett1996mixed,coffman2000distributed}:
\begin{equation}
    \mathcal{C}_{AB}(\rho) = \text{max}(0, \mu_1 - \mu_2 - \mu_3 - \mu_4)\,,
    \label{cur}
\end{equation}
where $\rho$ is the quantum state of the two qubits A and B, and $\mu_j$ are the eigenvalues of the matrix $\Tilde{\rho} = \rho S \rho^* S$, with $S = \sigma_y \otimes \sigma_y$. Here, $\sigma_y$ is the conventional Pauli matrix in the y direction. The eigenvalues are defined in descending order $\mu_j > \mu_{j+1}$. If $\mathcal{C}_{AB}(\rho)=0$, the state has no entanglement (separable state), and if $\mathcal{C}_{AB}(\rho)=1$, the system is maximally entangled. In Fig.~\ref{FigEntanglement}, we show the evolution of the concurrence between the donor qubits $\mathcal{C}_{12}$ as well as the concurrence between the acceptor qubits $\mathcal{C}_{34}$ for the two monomer case, described in section \ref{two_monomer_sec}.
}

\crr{
Initially, since the system is prepared in the state $\ket{T_D}$, which is maximally entangled, the concurrence is maximal, i.e., $\mathcal{C}_{12}=1$. As time elapses, the two qubits in the donor site lose all entanglement as they reach the separable state $\ket{\!\downarrow \downarrow}$ when the excitation has been transferred to the acceptor. In panel (a), where the system is set at a resonance ($\epsilon = 3 \omega$), and the temperature is low ($\bar{n}=0.01$), the steady state of the system is the electronic state $\vert T_A\rangle $, which is also a maximally entangled state of the two qubits, this is visible in the figure: $\mathcal{C}_{34} \rightarrow 1$ at long times.

In the ideal conditions of panel (a), the state of the phonons is the displaced vacuum state, so the steady state of the total system is the pure state $\ket{\psi_{T_D,0}}$. Consequently, $\rho^2 \rightarrow 1$ at long times, as reported in the plot. If we increase $\bar{n}$, as in panel (b), the steady state can be a mixed state of multiple phonon states, as shown by $\rho^2 <1$. Nonetheless, the electronic state steady state can still be purely $\ket{T_D}$, and in that case, all the initial entanglement between qubits in the donor site is still transferred to the acceptor site.

In panel (c), we move away from resonance by setting $\epsilon = 0.26 \omega$, such that the initial state $\ket{T_D}$ can weakly couple to $\ket{S_A}$. In that case, the electronic steady state is a mixture of $\ket{T_A}$ and $\ket{S_A}$, consequently, not all the initial entanglement is transferred to the acceptor, namely, $\mathcal{C}_{34} < 1$ in the steady state. 

When the donor and acceptor sites contain more than two qubits (see section \ref{sec_trimers}), other entanglement measures different from concurrence would be needed to characterize the entanglement dynamics of the mixed multi-qubit states. However, we expect similar entanglement dynamics. For instance, for the parameters chosen here, if the system is initialized in $\ket{\mathcal{E}_D^1}$, ideal conditions would lead to transfer into the state $\ket{\mathcal{E}_A^1}$ (see Fig.~\ref{Fig5}(b)). This state has a high overlap with state $\ket{W_A}$, as shown in Appendix F, which implies that the steady state of the system has a large entanglement. For instance, if we consider the donor/acceptor to have $N$ qubits, with $N$ being even, and consider the bipartition of $N/2$ qubits in each subsystem, the bipartite entanglement that can be transferred from the donor to the acceptor grows approximately as $S \propto \log_2(N) + C$ \cite{Dur2000} when the system is initialized in $\ket{\mathcal{E}_D^1}$.

}

\section{Configurations with static disorder} \label{app_disorder}

When considering the effects of static disorder, we defined the parameter sets $p_1$, $p_2$, $p_3$, and $p_4$ signaled with markers in Fig.~\ref{Fig3}(a). In Table~\ref{table:parameter sets}, we report the values of all parameters for each of these parameter sets.

\begin{table}[t!]
\centering
\begin{tabular}{lll}
\toprule
Label &  $\gamma/\omega$ & $p$ \\
\midrule
$p_1$ & 0.039552 & 1.0\\
$p_2$ & 0.010506 & 1.0\\
$p_3$ & 0.111707 & 1.0\\
$p_4$ & 0.030263 & 1.5\\
\bottomrule
\end{tabular}
\caption{Parameter sets used for the disorder analysis. All sets consider $J=0.3\omega$, $\epsilon = 3\omega$, $\bar{n}=0.01$, and $g=\omega$.}
\label{table:parameter sets}
\end{table}

\begin{figure}[t!]
\includegraphics[width=0.45\textwidth]{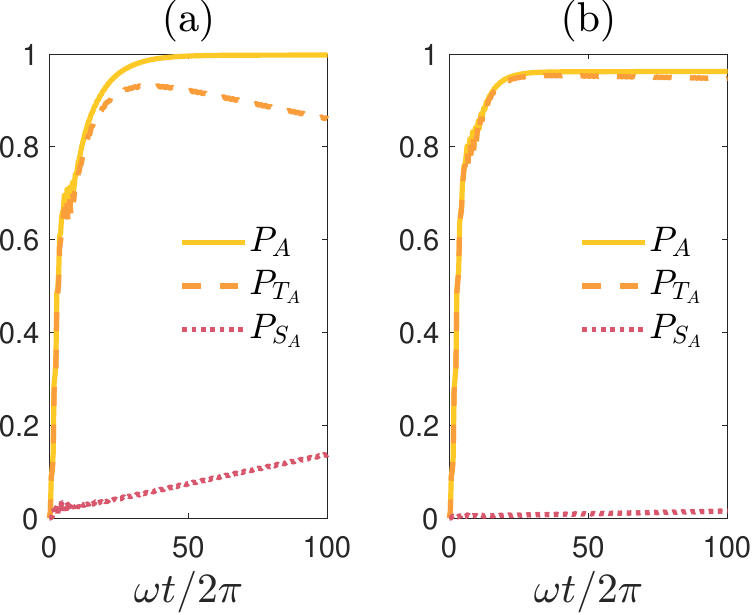}
\caption{{\bf Dynamics of specific disorder realizations:} Time evolution of the electronic-state populations for configurations with disordered parameters $g_j$. The yellow solid line, orange dashed line, and pink dotted line represent the population in the acceptor site, the triplet acceptor state, and the singlet acceptor state, respectively. %
{\bf(a)} The disordered parameters are given by $g_1/\omega = 0.9883$, $g_2/\omega=0.9906$, $g_3/\omega = -0.9638$, and $g_4/\omega = -1.1056$.
{\bf(b)} The disordered parameters are given by $g_1/\omega = 1.0923$, $g_2/\omega=0.9405$, $g_3/\omega = -1.0220$, and $g_4/\omega = -0.9806$.} 
\label{Fig2supp}
\end{figure}

As discussed in the main text, static disorder generally reduces the transfer rate, however, the effect of disorder in the dynamics can vary depending on the specific disorder realization. In Fig.~\ref{Fig2supp}, we consider two particular disordered configurations to illustrate this point.

The dynamics reported in panel (a) show that the disorder can cause other processes to become resonant. In this case, we see a coupling between the two acceptor states, increasing the population of the singlet state $\vert S_A\rangle$. At the perturbative level and in the absence of disorder, there is no coupling between these two states, however, the presence of disorder, even for small $J$, can modify this picture as $\vert T_A \rangle$ and $\vert S_A \rangle$ are not eigenstates of the unperturbed Hamiltonian anymore. However, we should note that this oscillation in the acceptor states might not significantly affect the transfer rate as the population of the donor decreases rapidly and remains close to zero.

 For the disorder realization in Fig.~\ref{Fig2supp}(b), on the other hand, we see that the population transfer to the acceptor is incomplete, namely, some population always remains in the donor. As we discussed in section~\ref{sec_temp}, a higher donor population in the steady state can decrease the value of the transfer rate in Eq.~\eqref{transferrate}.

\section{Higher complexity monomers and more monomers}
\label{app_monomer}

We start by considering the case of two monomers with three qubits per monomer ($L=3$) that we analyzed in section~\ref{sec_trimers}. The unperturbed Hamiltonian in the donor site is given by:

\begin{eqnarray} \label{H01D}
H_{0,D} &=& \frac{\epsilon}{2} (\vert 1 \rangle \langle 1 \vert + \vert 2 \rangle \langle 2 \vert + \vert 3 \rangle \langle 3) + \omega a^{\dagger}a \nonumber \\
&+& J\left(\left(\vert 1 \rangle \langle 2 \vert + \vert 2 \rangle \langle 3 \vert + \frac{1}{2}\vert 1 \rangle \langle 2 \vert \right) + {\rm h.c.}\right) \nonumber \\
&+& \frac{g}{2} (\vert 1 \rangle \langle 1 \vert + \vert 2 \rangle \langle 2 \vert + \vert 3 \rangle \langle 3\vert)\,,
\end{eqnarray}

where we have kept $p=1$. Diagonalization of Eq.~\eqref{H01D} yields the three eigenstates in Eq.~\eqref{epsiloeigenstates}, where $\mathcal{N}_1=\frac{1}{2} \sqrt{\frac{1}{2} \left(33-\sqrt{33}\right)}$ and $\mathcal{N}_2= \frac{1}{2} \sqrt{\frac{1}{2} \left(\sqrt{33}+33\right)}$. The corresponding eigenergies are given by:

\begin{eqnarray} \label{eigenenergiestrimer}
&&E_{\mathcal{E}_D^1}= \frac{1}{4} \left(\sqrt{33} J+J+4 \epsilon \right),\nonumber \\ &&E_{\mathcal{E}_D^2}= \frac{1}{4} \left(-\sqrt{33} J+J+4 \epsilon \right),\nonumber \\
&&E_{\mathcal{E}_D^3}= \epsilon -\frac{J}{2}\,.
\end{eqnarray}

$\ket{\mathcal{E}_D^1}$ is the state with the largest overlap with the $W$ state $\ket{W_D}$ and also the state with the highest eigenenergy. Similarly, equivalent results for states $\vert \mathcal{E}^{1,2,3}_A \rangle$ can be found for the acceptor sites in terms of states $\vert 4 \rangle$, $\vert 5 \rangle$, and $\vert 6 \rangle$. The eigenenergies of these states are identical to those in Eq.~\eqref{eigenenergiestrimer} but with negative on-site energy $\epsilon \rightarrow -\epsilon$. $\ket{\mathcal{E}_A^1}$ also corresponds to the acceptor state with the highest energy and largest overlap with $\vert W_A \rangle$. 

The perturbation Hamiltonian containing all the inter-monomer couplings can be written as:

\begin{equation} \label{perturbepsilon}
H_J = \sum_{A,B} J_{AB} \vert A \rangle \langle B \vert + {\rm h.c.}\,,
\end{equation}

where $A = \mathcal{E}_A^1, \mathcal{E}_A^2, \mathcal{E}_A^3$ and $B = \mathcal{E}_D^1, \mathcal{E}_D^2, \mathcal{E}_D^3$. In Table~\ref{table:JAB}, we report all the values of $J_{AB}$, where it can be noted that the process where the population is transferred between $\vert \mathcal{E}^1_D \rangle$ and $\vert \mathcal{E}^1_A \rangle$ is the fastest. Moreover, if the system is initially prepared in another state in the donor ($\vert \mathcal{E}_D^2 \rangle$ or $\vert \mathcal{E}_D^3 \rangle$), transfer to the $\vert \mathcal{E}_A^1 \rangle$ state is always favored at least from the perspective of having a stronger coupling $J_{AB}$ (see fourth and seventh rows in Table~\ref{table:JAB}).

\begin{table}[h!]
\centering
\begin{tabular}{ccc}
\toprule
Donor state & Acceptor state & $\vert J_{AB}/J \vert$ \\
\midrule
$\mathcal{E}^1$ & $\mathcal{E}^1$ & 0.6299\\
$\mathcal{E}^1$& $\mathcal{E}^2$ & 0.0553\\
$\mathcal{E}^1$ & $\mathcal{E}^3$ & 0.1446\\
$\mathcal{E}^2$ & $\mathcal{E}^1$ & 0.0786\\
$\mathcal{E}^2$& $\mathcal{E}^2$ & 0.0082\\
$\mathcal{E}^2$ & $\mathcal{E}^3$ & 0.0250\\
$\mathcal{E}^3$ & $\mathcal{E}^1$ & 0.0849\\
$\mathcal{E}^3$& $\mathcal{E}^2$ & 0.0103\\
$\mathcal{E}^3$ & $\mathcal{E}^3$ & 0.0381\\
\bottomrule
\end{tabular}
\caption{Inter-monomer couplings $J_{AB}$ in the basis of the eigenstates of Eq.~\eqref{H01D}.}
\label{table:JAB}
\end{table}

In the case of transfer from the initial $\vert W_D \rangle$ state (see Fig.~\ref{Fig5}(a)-(b)), the transfer rate is very similar to the case of initial $\vert \mathcal{E}_D^1 \rangle$ for two reasons: (i) the hierarchy of the $J_{AB}$ favors the transfer from any donor state to the $\vert \mathcal{E}_A^1 \rangle$ state, as discussed above, and (ii) these two initial states have a very large overlap $\vert \langle \mathcal{E}_D^1 \vert W_D \rangle \vert^2 = 0.993$. This overlap decreases as $L$ increases, but remains very large, for instance $\vert \langle \mathcal{E}_D^1 \vert W_D \rangle \vert^2 = 0.984$ for $L=5$.

The transfer from $\vert \mathcal{E}^1_D \rangle$ to $\vert \mathcal{E}^1_A \rangle$ remains the fastest process also for larger $L$. As an example, we consider the case $L=4$ with donor eigenstates:

\begin{eqnarray} \label{eigenstatesL4}
    &&\vert \mathcal{E}^m_D \rangle = \frac{1}{\mathcal{N}_m}\left(\vert 1 \rangle +\nu_{m,2} \vert 2 \rangle + \nu_{m,3}\vert 3 \rangle + \vert 4 \rangle\right), \nonumber \\
    &&\nu_{1,2}=\nu_{1,3}=\frac{1}{9}(2+\sqrt{85}), \nonumber \\
    &&\nu_{2,2}=\nu_{2,3}=\frac{1}{9}(2-\sqrt{85}), \nonumber \\
    &&\nu_{3,2}=-\nu_{3,3}=\frac{1}{3}(2-\sqrt{13}), \nonumber \\
    &&\nu_{4,2}=-\nu_{4,3}=\frac{1}{9}(2+\sqrt{13})\,.
\end{eqnarray}

Where the normalization constants are defined by $\mathcal{N}_m = \sqrt{2 + \nu_{m,2}^2 + \nu_{m,3}^2}$. Consistent with our notation, $\ket{\mathcal{E}_D^{1}}$ is the state with largest overlap with the $W$-state ($\vert \langle \mathcal{E}_D^1 \vert W_D \rangle \vert^2 = 0.988$) and with the largest eigenenergy $E_{\mathcal{E}_D^1} = \frac{1}{6} \left(\sqrt{85}+4\right) J+\epsilon$. In Table~\ref{table:JABL4}, we report the magnitude of all inter-monomer couplings when the perturbation Hamiltonian is written in the form of Eq.~\eqref{perturbepsilon}. Apart from the transfer from $\vert \mathcal{E}^1_D \rangle$ to $\vert \mathcal{E}^1_A \rangle$ being the fastest, we also note that the transfer to state $\vert \mathcal{E}_A^1 \rangle$ is always favored over other processes regardless of the choice of initial state on this basis.

\begin{table}[h!]
\centering
\begin{tabular}{ccc}
\toprule
Donor state & Acceptor state & $\vert J_{AB}/J \vert$ \\
\midrule
$\mathcal{E}^1$ & $\mathcal{E}^1$ & 0.7083\\
$\mathcal{E}^1$& $\mathcal{E}^2$ & 0.1280\\
$\mathcal{E}^1$ & $\mathcal{E}^3$ & 0.2036\\
$\mathcal{E}^1$ & $\mathcal{E}^4$ & 0.0211\\
$\mathcal{E}^2$ & $\mathcal{E}^1$ & 0.0823\\
$\mathcal{E}^2$& $\mathcal{E}^2$ & 0.0186\\
$\mathcal{E}^2$ & $\mathcal{E}^3$ & 0.0342\\
$\mathcal{E}^2$ & $\mathcal{E}^4$ & 0.0039\\
$\mathcal{E}^3$ & $\mathcal{E}^1$ & 0.1025\\
$\mathcal{E}^3$& $\mathcal{E}^2$ & 0.0267\\
$\mathcal{E}^3$ & $\mathcal{E}^3$ & 0.0576\\
$\mathcal{E}^3$ & $\mathcal{E}^4$ & 0.0065\\
$\mathcal{E}^4$ & $\mathcal{E}^1$ & 0.0371\\
$\mathcal{E}^4$& $\mathcal{E}^2$ & 0.0106\\
$\mathcal{E}^4$ & $\mathcal{E}^3$ & 0.0228\\
$\mathcal{E}^4$ & $\mathcal{E}^4$ & 0.0027\\
\bottomrule
\end{tabular}
\caption{Inter-monomer couplings $J_{AB}$ in the basis of the eigenstates of the unperturbed Hamiltonian (see Eq.~\eqref{perturbepsilon}) for $L=4$.}
\label{table:JABL4}
\end{table}

\section{Experimental realization of the vibrational and vibronic coupling terms}
\label{app_crosstalk}

Here we outline in detail the derivation of the trapped-ion Hamiltonian corresponding to the vibrational and vibronic coupling terms in Eq.~\eqref{eq_H_model} via the proposed experimental scheme in Sec. \ref{sec_ions1} and the discussion on the effect of the unwanted interactions arising from this scheme.

Consider the application of individual-addressing Raman beams with the same wavevector difference $\vec{k}_a$ along a radial direction, phase differences $\phi_i$, and a common beatnote frequency $\omega_L$ on the qubit ions. The trapped-ion Hamiltonian is ($\hbar = 1$):
\begin{align}
    H = & \sum_i\frac{\omega_0}{2}\sigma_i^{z} +\sum_\nu\omega_\nu a^\dagger_{\nu} a_{\nu} \nonumber \\ + & \sum_{i}\frac{\Omega_i}{2}\left[e^{\sum_\nu i\eta_{i\nu}\left(a_\nu +a^\dagger_\nu \right)-i\omega_L t-i\phi_i}\sigma_i^{+} + \text{h.c.}\right],
    \label{eq_Hfull}
\end{align}
where $\omega_0$ is the frequency difference between the internal states of the qubit ions, $\omega_\nu$ is the $\nu$-th collective mode frequency of the chain associated with the raising (lowering) operator $a^\dagger_\nu (a_\nu)$, $\Omega_i$ is the Rabi coupling strength to ion $i$, and $\eta_{i\nu} = k_a\sqrt{1/2m\omega_\nu}b_{i\nu}$ is the Lamb-Dicke parameter with the qubit mass $m$. $b_{i\nu}$ is the normalized motional eigenvector for ion $i$ in the collective mode $\nu$, where the lowest and highest frequency modes $\omega_1$ and $\omega_N$ are known as the zig-zag and center-of-mass modes, respectively.
\par Under the transformation with respect to $\sum\limits_i\frac{\omega_0}{2}\sigma_i^{z} + \sum\limits_{\nu'}\omega_{\nu'}a^\dagger_{\nu'}a_{\nu'} + \mu a_k^\dagger a_k$, Eq.~\eqref{eq_Hfull} becomes a resonant interaction Hamiltonian that rotates at $\mu = \omega_L - \omega_0 \equiv \omega_k+\delta_k$, where $\delta_k$ is the detuning from the $k$-th motional mode, and $\forall\nu' = \forall\nu \backslash \{k\}$ \cite{Schneider_2012}. For $\mu+\omega_k\gg|\mu-\omega_k|=\delta_k$, we neglect the terms that rotate at $\mu+\omega_k$ under rotating-wave approximation (RWA), which results in the transformed Hamiltonian:
\begin{align}
    H^\text{res} = & \sum_i\frac{\Omega_i}{2}\left[e^{\sum\limits_{\nu'} i\eta_{i\nu'}\left(a_{\nu'} e^{-i \omega_{\nu'} t}+a_{\nu'}^\dagger e^{i \omega_{\nu'} t}\right)} \right. \nonumber \\ & \left. e^{i\eta_{ik}\left(a_k e^{-i \mu t}+a_k^\dagger e^{i \mu t}\right)}e^{-i(\omega_L - \omega_0) t-i\phi_i}\sigma_i^{+}\right. \nonumber \\ &+ \left. \text{h.c.}\right] - \delta_k a^\dagger_k a_k \equiv \; H_{I}^\text{res} + H_{\delta}^\text{res}.
    \label{eq_interactionH0}
\end{align}
\par The second term above, $H_{\delta}^\text{res}=-\delta_k a_k^\dagger a_k$, corresponds to the vibrational energy term in Eq.~\eqref{eq_H_model} with $-\delta_k \mapsto \omega$ and $a_k^\dagger (a_k) \mapsto a^\dagger (a)$. To obtain the vibronic coupling term, we apply individual-addressing Raman beams with two symmetric beatnote frequency tones $\omega_r = \omega_0-\mu$ with controlled phase $\phi_i^{r}$ and Rabi frequency $\Omega_i^{r}$ and $\omega_b = \omega_0+\mu$ with controlled phase $\phi_i^{b}$ and Rabi frequency $\Omega_i^{b}$. Hence, the interaction Hamiltonian becomes:
\begin{align}
    H_{I}^\text{res} = & \sum_{l=r,b}\sum_i\frac{\Omega_i^{l}}{2}\left[e^{\sum\limits_{\nu'} i\eta_{i\nu'}\left(a_{\nu'} e^{-i \omega_{\nu'} t}+a_{\nu'}^\dagger e^{i \omega_{\nu'} t}\right)} \right. \nonumber \\ & \left. e^{i\eta_{ik}\left(a_k e^{-i \mu t}+a_k^\dagger e^{i \mu t}\right)}e^{-i(\omega_l - \omega_0) t-i\phi_i^{l}}\sigma_i^{+} \right. \nonumber \\ & \left. + \;\text{h.c.}\right].
    \label{eq_interactionHsp}
\end{align}
 
 In the Lamb-Dicke regime, where $\eta_{i,\nu}\sqrt{\braket{\left(a_\nu+a^\dagger_\nu\right)^2}}\ll1$, we can expand the interaction Hamiltonian with respect to $\eta_{i\nu}$ to the first order. For $\Omega_i^{r} = \Omega_i^{b} \equiv \Omega_i^{\text{sp}}$, we obtain:
\begin{align}
    H_I^\text{res} = & \medspace \sum_i\Omega_i^\text{sp}\cos{(\mu t + \phi_i^{m})}\sigma_{\phi_i^{s}+\pi/2} \nonumber \\ + &\sum_i\frac{\eta_{ik}\Omega_i^\text{sp}}{2}\left[a_k \left(e^{i\phi_i^{m}}+e^{-i2\mu t-i\phi_i^{m}}\right) \right. \nonumber  \\ + & \left. a_k^\dagger \left(e^{-i\phi_i^{m}}+e^{i2\mu t+i\phi_i^{m}}\right)\right]\sigma_{\phi_i^{s}} \nonumber  \\ + & \sum_{\nu'}\sum_{i}\eta_{i\nu'}\Omega_i^\text{sp}\left(a_{\nu'} e^{-i\omega_{\nu'}t} + a_{\nu'}^\dagger e^{ i\omega_{\nu'}t}\right) \nonumber  \\ & \cos{(\mu t + \phi_i^{m})}\sigma_{\phi_i^{s}},
\end{align}
where the motional and spin phases are denoted by $\phi_i^{m} \equiv \frac{\phi_i^{b} - \phi_i^{r}}{2}$ and $\phi_i^{s} \equiv \frac{\phi_i^{b} + \phi_i^{r}}{2}-\frac{\pi}{2}$, respectively, and we define the spin operator $\sigma_{\phi_i^{s}}\equiv\cos(\phi_i^{s})\sigma_i^{x}+\sin(\phi_i^{s})\sigma_i^{y}$ \cite{monroe2021programmable}. We note that $\sigma_i^{y} \rightarrow \sigma_i^{z}$ in the driven rotating lab frame.

\par Since the interaction Hamiltonian is time-dependent, we can approximate the time-ordering evolution operator, $U(t)=T\left[\exp{\left(-i\int_{0}^t dt_1 H_I^\text{res}(t_1)\right)}\right]$, with the Magnus expansion and obtain $U(t)\approx\exp{\left(-iH_I^\text{eff}t\right)}$ with the effective Hamiltonian:
\begin{align}
    H_I^\text{eff} = & \medspace \sum_i\frac{\eta_{ik}\Omega_i^\text{sp}}{2}\left(a_k e^{i\phi_i^{m}} + a_k^\dagger e^{ -i\phi_i^{m}}\right)\sigma_{\phi_i^{s}} \nonumber\\ + & \medspace \sum_{\nu'}\sum_{i<j}\frac{\eta_{i\nu'}\eta_{j\nu'}\omega_{\nu'}}{\mu^2-\omega_{\nu'}^2}\Omega_i^{\text{sp}}\Omega_j^{\text{sp}}\cos{(\phi_i^{m} - \phi_j^{m})} \nonumber \\ & \sigma_{\phi_i^{s}}\sigma_{\phi_j^{s}},
    \label{H_spfull}
\end{align}
for when $\delta_{\nu'}\equiv\mu-\omega_{\nu'}\gg\eta_{i\nu'}\Omega_i^\text{sp}$. Here, the first term is the desired vibronic coupling term in Eq.~\eqref{eq_H_model} with $\eta_{ik}\Omega_i^\text{sp} \mapsto g_i$ and an appropriate choice of optical phases, while the second term accounts for off-resonant couplings to other motional modes, which leads to an unwanted two-body electronic coupling term:
\begin{align}
    H_{ij}^\text{cr} \equiv & \medspace \sum_{\nu'}\sum_{i<j}\frac{\eta_{i\nu'}\eta_{j\nu'}\omega_{\nu'}}{\mu^2-\omega_{\nu'}^2}\Omega_i^{\text{sp}}\Omega_j^{\text{sp}}\cos{(\phi_i^{m} - \phi_j^{m})} \nonumber \\ & \sigma_{\phi_i^{s}}\sigma_{\phi_j^{s}}.
\end{align}
In the rotated frame that the quantum simulation of our model is performed, the interaction becomes
\begin{equation}
    H_{ij}^\text{cr} = \sum_{i<j}J_{ij}^\text{cr} \sigma_{i}^{z}\sigma_{j}^{z},
\end{equation}
where $J_{ij}^\text{cr}\equiv\sum\limits_{\substack{\nu'}}\frac{\eta_{i\nu'}\eta_{j\nu'}\omega_{\nu'}}{\mu^2-\omega_{\nu'}^2}\Omega_i^{\text{sp}}\Omega_j^{\text{sp}}\cos{(\phi_i^{m} - \phi_j^{m})}$. Therefore, in the relevant electronic manifold, where one of the ions is in the spin-up state representing the location of the electronic excitation ($\{\vert i \rangle\}$), the unwanted two-body electronic coupling Hamiltonian corresponds to a state-dependent shift in energy for the relevant spin configurations:
\begin{eqnarray}
\langle k \vert H_{ij}^\text{cr} \vert k \rangle &=& \sum\limits_{i<j}f(k,i,j)J_{ij}^\text{cr}, \\
f(k,i,j)&=& \begin{cases}
-1, & \text{if} \; i=k \; \text{or} \; j = k,\\
+1, & \text{otherwise.}
\end{cases}.
\end{eqnarray}


Similarly, in any basis spanned by $\{\vert i \rangle\}$, the interaction Hamiltonian also leads to state-dependent energy shifts. For instance, in the $\{\vert T_D \rangle,\vert S_D \rangle,\vert T_A \rangle,\vert S_A \rangle\}$ basis used to study the two-monomer model in Sec. \ref{two_monomer_sec}, we obtain:
\begin{eqnarray}
\langle T_D \vert H_{ij}^\text{cr} \vert T_D \rangle &=& - J_{12}^\text{cr} + J_{34}^\text{cr} \nonumber \\
\langle S_D \vert H_{ij}^\text{cr} \vert S_D \rangle &=& - J_{13}^\text{cr} - J_{14}^\text{cr} + J_{23}^\text{cr} + J_{24}^\text{cr}  \nonumber \\
\langle T_A \vert H_{ij}^\text{cr} \vert T_A \rangle &=& + J_{12}^\text{cr} - J_{34}^\text{cr} \nonumber \\
\langle S_A \vert H_{ij}^\text{cr} \vert S_A \rangle &=& - J_{13}^\text{cr} + J_{14}^\text{cr} - J_{23}^\text{cr} + J_{24}^\text{cr}. \nonumber \\
\end{eqnarray}
Hence, this two-body electronic coupling crosstalk term should only induce changes in the energy landscape and, in turn, affect the excitation transfer resonance conditions while preserving the physical dynamics of the system unless an appropriate energy compensation is introduced.

\end{document}